\documentclass[a4paper,11pt]{article}
\pdfoutput=1

\usepackage[utf8]{inputenc}
\usepackage[margin=1in]{geometry}
\usepackage{mathtools}
\usepackage{amsfonts}
\usepackage{amssymb}
\usepackage{amsmath}
\usepackage{tikz}
\usepackage{tikz-cd}
\usetikzlibrary{decorations.pathreplacing, decorations.markings}
\usepackage{hyperref}
\usepackage{xspace}
\usepackage{authblk}
\usepackage{todonotes}

\newcommand{\df}{D6$_5$\xspace}
\newcommand{\ds}{D6$_7$\xspace}

\title{M-theory geometries from five-brane webs, seven-branes, and T-branes} 
\author{Andr\'es Collinucci}
\affil{Physique Théorique et Mathématique and International Solvay Institutes \\
Université Libre de Bruxelles, C.P. 231, 1050 Brussels, Belgium}

\begin{document}
\maketitle

\begin{abstract}
    We track the chain of dualities relating five-brane webs in Type IIB to
    M-theory on noncompact Calabi--Yau threefolds, and follow the effect of
    adding $(p,q)$ seven-branes. T-dualizing along the seven-brane, a junction
    of D5-branes ending on a D7-brane becomes a single smooth D6-brane wrapping a
    smooth holomorphic curve, which uplifts directly to an M-theory geometry.
    Treating the branes as coherent sheaves (tachyon condensation), we obtain a
    spectral-curve dictionary that maps non-Abelian (T-brane) partition data to
    explicit complex-structure deformations of the threefold. As applications we
    (i) give a physical derivation of these deformations, (ii) exhibit a simple
    geometric prototype of an s-rule violation, and (iii) link T-brane data to
    geometry.
\end{abstract}

\tableofcontents

\section{Introduction}
There is a well-known string theory construction of 5d SCFT's, that has a two-fold incarnation: A web of IIB $(p,q)$ 5-branes on one side, and M-theory on a non-compact, toric Calabi-Yau singularity on the other side. 

The goal of this paper is to enrich this correspondence, by including the effects of 7-branes on the IIB side, and track their geometric repercussions on the M-theory side. As expected, we will see complex structure deformations that break toricity.

Several works have been written to tackle the problem, see \cite{Arias-Tamargo:2024fjt, CarrenoBolla:2025rkv, CarrenoBolla:2024fxy, Cremonesi:2023psg}. Most vexing is the 5d version of the \emph{s-rule}, which selects consistent brane-web configurations, which has remained elusive and subject to debate in the community. From the mathematical side, a new framework has been introduced by Alexeev, Arg\"uz and Bousseau \cite{Alexeev:2024bko} that addresses this by exploiting local mirror symmetry, and mapping the complex structure deformations induced by movements of the sevenbranes into blow-ups of the mirror threefold. The methodology is powerful and promising, but quite involved, and hence not readily accessible to the physics community.

In this paper, we will take a different approach. We will chase the dualities that take us from IIB to M-theory, and track the effect of such 7-branes. In the simplest cases, D7-D5 junctions dualize to D6-branes wrapping smooth Riemann surfaces, which then readily uplift to Taub-NUT type geometries in M-theory. We will mimic the logic of \cite{Witten:1997sc} for tracking how D4-NS5 junctions lift to smooth M5-Riemann surfaces. Even though the context is quite different, the mathematics are the same: In Witten's setup, an D4/NS5 setup maps to an M5-brane wrapping a Riemann surface that accounts for the smoothing of the D4/NS5-junctions. In our setups, a D5/D7 setup, with its junctions smoothed out, T-dualizes to a smooth D6-brane wrapping a Riemann surface. Armed with a clear dictionary between D5's ending on D7-branes in IIB, and smooth D6-Riemann surfaces in IIA, we will be able to write down the uplifted M-theory geometries.

The main results of this paper are the following:
\begin{enumerate}
    \item We give a physical derivation of the complex structure deformations of the M-theory 3-fold induced by 5-branes ending on a 7-brane.
    \item We provide a simple prototype example of a violation of the s-rule, and its geometrical counterpart. We will see that inconsistent brane setups lead to geometries with poles in their defining equations.
    \item By using the tachyon condensation picture of branes, i.e. treating them as coherent sheaves, we establish a clear link between \emph{T-brane} data and geometry.
\end{enumerate}
A few comments about these three points in brief. Point 1 is the building block. One can generalize the construction to general 5-brane/7-brane junctions even beyond the purely perturbative case by zooming in on the external legs of any given 5-brane web, and choosing the right $SL(2, \mathbb{Z})$-frame. The remaining challenge then is gluing such pieces back together into a consistent web.
Our results will even show us what a Hanany-Witten brane creation/destruction in IIB looks like on the M-theory side.

Point 2 has been addressed from many viewpoints in the literature. Here we provide the following observation in a prototype case, two D5-branes suspended between an NS5 and a D7. In IIB, a Hanany-Witten transition is expected to trigger the creation of an \emph{anti}-D5 segment, incompatible with the ambient supersymmetry. What we find on our side is the appearance of poles in the defining geometry. We will comment on this in the paper.

Point 3 introduces a piece of technology to explicitly map partition data, meaning, how many 5-branes end on each 7-brane in the setup, into a non-Abelian adjoint field whose spectral data captures the M-theory geometry. This was studied from a different point of view in \cite{Bourget:2023wlb}, but here we corroborate it with a derivation.

\section{M-theory/IIB duality: Recap}
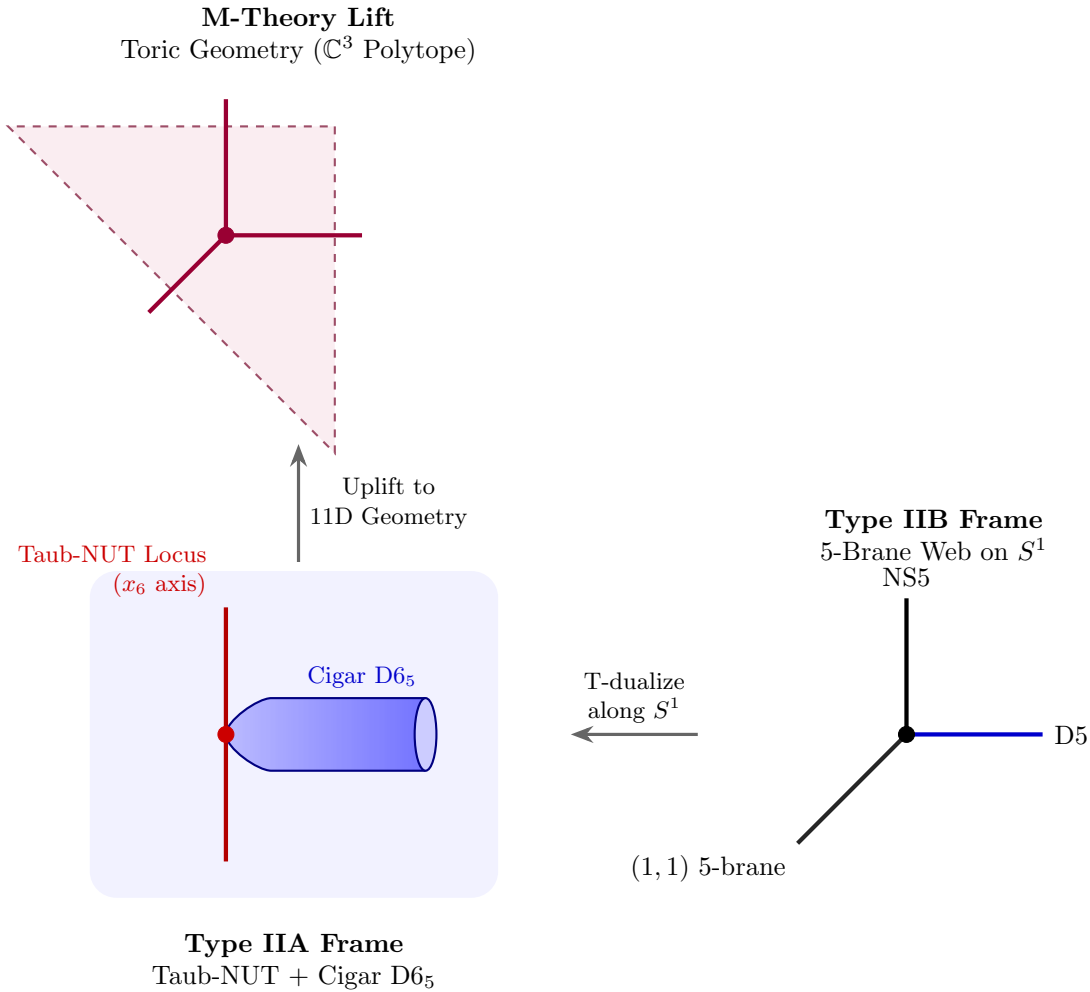
\begin{figure}[h]
\begin{tikzpicture}[scale=1.2, every node/.style={font=\small}]

    % ==========================================
    % LEFT PANEL (Bottom-Left): Type IIA Situation
    % ==========================================
    \begin{scope}[shift={(0, 0)}]
        % Shaded background for the local patch
        \fill[blue!5, rounded corners=10pt] (-1.5, -1.8) rectangle (3.0, 1.8);
        
        % TN Locus extended along x_6
        \draw[ultra thick, red!70!black] (0, -1.4) -- (0, 1.4);
        
        % Shifted label slightly left
        \node[red!80!black, above left, font=\footnotesize, align=right] at (-0.1, 1.4) {Taub-NUT Locus \\ ($x_6$ axis)};
        
        % The Cigar-like D6_5 Brane terminating exactly on the x_6 axis
        \shade[left color=blue!30, right color=blue!65, opacity=0.85]
            (0,0) to[out=90, in=180, looseness=0.6] (0.5, 0.4) -- (2.2, 0.4)
            arc(90:-90:0.12 and 0.4) -- (0.5, -0.4)
            to[out=180, in=-90, looseness=0.6] (0,0) -- cycle;
            
        \draw[blue!50!black, thick] (0,0) to[out=90, in=180, looseness=0.6] (0.5, 0.4) -- (2.2, 0.4);
        \draw[blue!50!black, thick] (0,0) to[out=-90, in=180, looseness=0.6] (0.5, -0.4) -- (2.2, -0.4);
        \filldraw[fill=blue!20, draw=blue!50!black, thick] (2.2, 0) ellipse (0.12 and 0.4);
        
        % The explicit zero-point center
        \filldraw[red!80!black] (0,0) circle (2.5pt);
        
        \node[blue!80!black, above, font=\footnotesize] at (1.5, 0.4) {Cigar D6$_5$};
        \node[align=center] at (0.75, -2.5) {\textbf{Type IIA Frame}\\ Taub-NUT + Cigar D6$_5$};
    \end{scope}

    % ==========================================
    % RIGHT PANEL (Bottom-Right): IIB 5-Brane Web
    % ==========================================
    % Horizontal shift set to 7.5 to create comfortable separation
    \begin{scope}[shift={(7.5, 0)}]
        \node[align=center] at (0.3, 2.2) {\textbf{Type IIB Frame}\\ 5-Brane Web on $S^1$};
        
        \draw[ultra thick, blue!80!black] (0,0) -- (1.5,0) node[right, black] {D5};
        \draw[ultra thick, black] (0,0) -- (0,1.5) node[above] {NS5};
        \draw[ultra thick, black!70!gray] (0,0) -- (-1.2,-1.2) node[below left, black] {$(1,1)$ 5-brane};
        
        \filldraw[black] (0,0) circle (2.5pt);
    \end{scope}

    % ==========================================
    % TOP PANEL (Top-Left): M-Theory Uplift
    % ==========================================
    \begin{scope}[shift={(0, 5.5)}]
        \node[align=center] at (0.8, 2.2) {\textbf{M-Theory Lift}\\ Toric Geometry ($\mathbb{C}^3$ Polytope)};
        
        % Centered Toric Polytope
        \filldraw[fill=purple!10, draw=purple!60!black, thick, dashed, opacity=0.7]
            (1.2, 1.2) -- (1.2, -2.4) -- (-2.4, 1.2) -- cycle;
        
        % The smooth geometry toric graph
        \draw[ultra thick, purple!80!black] (0,0) -- (1.5, 0); 
        \draw[ultra thick, purple!80!black] (0,0) -- (0, 1.5); 
        \draw[ultra thick, purple!80!black] (0,0) -- (-0.85, -0.85); 
        
        \filldraw[purple!80!black] (0,0) circle (2.5pt);
    \end{scope}

    % ==========================================
    % DUALITY TRANSITION ARROWS
    % ==========================================
    
    % Horizontal Arrow: IIB to IIA 
    \draw[-{Stealth[length=3mm]}, line width=1.2pt, gray!80!black] 
        (5.2, 0) -- (3.8, 0) 
        node[midway, above, align=center, black, font=\footnotesize] {T-dualize \\ along $S^1$};

    % Vertical Arrow: IIA to M-Theory
    \draw[-{Stealth[length=3mm]}, line width=1.2pt, gray!80!black] 
        (0.8, 1.9) -- (0.8, 3.2) 
        node[midway, right, align=center, black, font=\footnotesize] {Uplift to \\ 11D Geometry};

\end{tikzpicture} 
\caption{Chain of dualities realizing the 5-brane web/M-theory on CY threefolds correspondence. In this case, $\mathbb{C}^3$ is shown.}\label{fig:vafaleung}
\end{figure}

In this section, we briefly remind the reader of the setting, and the problem at hand. We refer to the seminal papers for more details: In \cite{Aharony:1997bh}, 5-brane webs are introduced; and in \cite{Benini:2009gi} sevenbranes are added to the picture.
The correspondence, sometimes referred to as Leung-Vafa duality \cite{Leung:1997tw}, establishes a dictionary between IIB string theory in the presence of a web of $(p, q)$ 5-branes and a noncompact toric Calabi-Yau threefold. The toric CY can be described by a polytope, which corresponds to the `ceiling' of its 3d fan, and the brane web can be read off this polytope by simply drawing the dual graph.

Two physical phenomena severely alter this 5-brane web/toric CY correspondence:
\begin{enumerate}
    \item Allowing $(p,q)$ 5-branes to terminate on $(p,q)$ 7-branes. One can recover the usual dictionary by sending the 7-branes to infinity, however, bringing them to within finite distance is expected to correspond to a complex structure deformation of the toric CY such that it is no longer toric.
    \item Instead of each $(p,q)$ 5-brane terminating on a single $(p,q)$ 7-brane, allowing for several 5-branes to end on a single 7-brane; or more generally, we allow for a non-trivial partition of the 5-branes as they terminate on the 7-branes. In \cite{Bourget:2023wlb}, it was claimed that T-brane data is required. T-branes are non-Abelian bound states of branes, first discussed in \cite{Gomez:2000zm} and then further developed in \cite{Cecotti:2010bp}.
\end{enumerate}
We can understand this duality as the following chain of dualities: Compactify IIB on a circle along the 7-branes transverse to the D5-branes, transverse to the NS5-branes. Then T-dualize to IIA. Now the NS5's have become pure Taub-NUT geometry, and the D5's and D7's have become D6-branes. Finally, uplift this to M-theory. Now the D6-branes become pure Taub-NUT geometry. The standard duality without 7-branes is illustrated in figure \ref{fig:vafaleung}. The semi-infinite D5 becomes a D6 in the shape of  a cigar, whose tip lies at the Taub-NUT center. Allowing, say, the D5 to terminate on a D7 will translate to altering the shape of that cigar, not topologically, but at the level of complex structure.

This chain of dualities  cannot handle mutually non-local branes. Hence, it but must be done patch-wise, and be glued back together.

\section{Setting up the problem}
\subsection{IIB setup}

We begin with the simplest situation: a collection of D5-branes ending on a single
D7-brane. In anticipation of T-dualizing to IIA, we place the system on a circle
$\theta := x_9$. The setup is as follows:
\begin{equation*}
    \begin{array}{ccccccccccc}
         &  0&1&2&3&4&5&6&7&8&9\\
         {\rm D}5&-&-&-&-&-&(-\infty,0]&&&&\\
         {\rm D}7&-&-&-&-&-& &&-&-&-\\
         S^1&&&&&&&&&&-
    \end{array}
\end{equation*}
Let $\phi_5$ be the real worldvolume scalar parametrizing the D7's position along $x_5$. We define the complex worldvolume coordinate $z := x_7 + i x_8$. The D5's span a semi-infinite
part of the $x_5$-axis and terminate at codimension-three loci on the D7-brane. Our aim here is to determine the worldvolume profile $\phi_5$ \emph{exactly}, without
passing through a two-dimensional effective description. However, one can take the quick route: Reducing the D7 worldvolume on the circle and solving a Poisson equation in the effective 2d theory. This will tell us that $\phi_5$ is a logarithm, however a potential tower of corrections will be absent by construction. We will instead solve the full problem on
$\mathbb{R}^2\times S^1$ and show that the logarithm is the \emph{complete}
perturbative answer, corrected only by an exponentially suppressed tower.

The endpoint of a D5-brane is a unit magnetic source for the D7 gauge field,
$\oint_{S^2} F = 2\pi$. Supersymmetry ties the physical transverse scalar $\phi_5$ to $F$ via
the Bogomolny equation, which explicitly includes the string length squared to balance dimensions:
\begin{equation}
    d\phi_5 = 2\pi\alpha' \ast_3 F\,.
\end{equation}
Acting with $d \ast$ yields a 3d Poisson equation strictly sourced by the monopole:
\begin{equation}\label{eq:laplace}
    \nabla^2_{(3d)} \phi_5 = 4\pi^2 \alpha'\,\delta^{(3)}(\vec{x}-\vec{x}_i)\,,
\end{equation}
with $\vec{x} = (x_7, x_8, \theta)$ and $\vec{x}_i$ the D5 endpoints. Decomposing this as 
\begin{equation}
\left(\nabla^2_{(2d)}+\partial_\theta^2\right) \phi_5 = 4 \pi^2 \alpha' \, \delta^{(2)}(z-z_i)\,\delta(\theta-\theta_0)\,,
\end{equation}
we can take a shortcut by decomposing in Fourier modes and keep just the zero-mode:
\begin{equation}
    \nabla^2_{(2d)} \phi^{(0)}_5(z) = \frac{4 \pi^2 \alpha'}{2 \pi R}\,\delta^{(2)}(z-z_i)\,.
\end{equation}
The natural solution is a logarithm $\phi_5 \sim \frac{\alpha'}{R}\,\log |z-z_0|$.
However, it is instructive to keep the full 3d equation on $\mathbb{R}^2\times S^1$. On the covering space the periodic Green's function is the sum of periodic image-charges:
\begin{equation}\label{eq:kernel}
    G(\vec{x}) = \sum_{n \in \mathbb{Z}}
    \frac{1}{4\pi\sqrt{|z-z_0|^2 + (\theta-\theta_0 - 2\pi n R)^2}}\,,
    \qquad \phi_5 = -\,4\pi^2 \alpha'\,G\,,
\end{equation}
with $(z_0,\theta_0)$ the chosen D5 endpoint. 
Poisson-resumming this expression into KK momentum modes yields:
\begin{equation}\label{eq:poisson}
    G = \frac{1}{4\pi^2 R}\,K_0\!\big(\mu|z-z_0|\big)\Big|_{\mu\to 0}
      + \frac{1}{4\pi^2 R}\sum_{m \neq 0}
        K_0\!\left(\frac{|m||z-z_0|}{R}\right) e^{i m (\theta-\theta_0)/R}\,,
\end{equation}
where $K_0$ is the zeroth modified Bessel function of the second kind. 
The two pieces separate cleanly. The $m=0$ term yields the expected solution upon first dimensionally reducing the equations of motion:
as the IR regulator $\mu\to 0$ it develops the expected logarithm,
$K_0(\mu\rho)\sim -\log(\mu\rho)-\gamma$. We absorb $\mu$, the Euler--Mascheroni
constant $\gamma$, and the integration constants into a physical reference scale
$\Lambda$. 

The $m\neq 0$ tower consists of massive Kaluza-Klein modes. Their radial behavior is governed by the modified Bessel function, which decays exponentially at large distances as $K_0(x) \sim \sqrt{\pi/2x}\,e^{-x}$. Therefore, these modes are highly suppressed for $|z-z_0| \gg R$. Restoring the source strength via \eqref{eq:kernel}, the full exact profile of a single source is:
\begin{equation}\label{eq:pertexact}
    \phi_5 = \frac{\alpha'}{R}\,\log\frac{|z-z_0|}{\Lambda}
    \;-\; \frac{\alpha'}{R}\sum_{m \neq 0}
      K_0\!\left(\frac{|m||z-z_0|}{R}\right) e^{i m (\theta-\theta_0)/R}\,.
\end{equation}
Hence, the 2D logarithm is the exact zero-mode profile. The full 3D geometry simply dresses this logarithm with a closed-form tower of heavy KK modes that smooth out the solution near the intersection.\footnote{We thank A. Tomasiello for suggesting to perform this check.}
. (This mathematical structure precisely parallels the Ooguri--Vafa metric \cite{Ooguri:1996me}, where the exact same periodic Green's function resolves a semi-flat logarithmic singularity via exponentially suppressed $K_0$ corrections). From now on, we will truncate to the `perturbative' logarithmic part.

Superposing $q_L$ sources pulling from the left and $q_R$ from the right, the
perturbative ($m=0$) profile is
\begin{equation} \label{reallogsolution}
    \phi_5 = \frac{\alpha'}{R} \sum_{i=1}^{q_L} \log\!\left(\frac{|z-z^{(-)}_i|}{\Lambda}\right)
           - \frac{\alpha'}{R} \sum_{j=1}^{q_R} \log\!\left(\frac{|z-z^{(+)}_j|}{\Lambda}\right)
           + \frac{\alpha'}{R} \log c_0\,,
\end{equation} where $\Lambda$ and $c_0$ are integration constants.  Physically, the D5-branes tug at the D7, dimpling it left or right. For left-right balanced configurations,
$Q_{\rm net} := q_L - q_R = 0$, the dependence on the arbitrary IR scale $\Lambda$
drops out, and $c_0$ fixed the position of the D7 in its flat region at large $z$.

\subsection{IIA T-dual setup}
Upon T-dualizing the $x_9$ circle, the D7-brane worldvolume gauge connection $A_9$ becomes a third transverse scalar, $\phi_9$, of the dual ${\rm D}6_F$-brane. To preserve supersymmetry, the resulting D6-brane must wrap a holomorphic curve in the target space. This BPS condition strictly enforces the Cauchy-Riemann equations on the transverse fluctuations, requiring $\phi_5$ and $\phi_9$ to pair\footnote{One could have \emph{a priori} paired $\phi_5$ with $\phi_6$. Indeed in absence of a D5, this would have dual to a hyper-K\"ahler rotation in the resulting M-theory geometry. However the presence of the D5 freezes $\phi_6$.} into an analytic function $\Phi = \phi_5 + i \phi_9$ of $z = x_7 + i x_8$. Since the Poisson equation fixes the real part to $\phi_5 \sim \frac{\alpha'}{R}\,\log|z|$, holomorphy uniquely demands that the real logarithm becomes a complex one $\Phi \sim \frac{\alpha'}{R}\,\log(z)$. After T-dualizing the circle to one of radius $\tilde R = \frac{
\alpha'}{R}$, the log now has the form
\[
\tilde \phi_5 \sim \tilde R\,\log(z-z_i)\,
\]
Hence, the unit of magnetic flux sourced by the original D5-brane endpoint translates to a monodromy of the dual $\tilde x_9$ coordinate. 

\begin{figure}
    \centering
    % ==========================================
    % LEFT PANEL: Intersecting (alpha = 0)
    % ==========================================
    \begin{minipage}{0.48\textwidth}
        \centering
        \begin{tikzpicture}[scale=1]
            % Axes
            \draw[->, thick, gray] (-3.5, -2.5) -- (3.5, -2.5) node[right] {$x_5$};
            \draw[->, thick, gray] (-3.5, -2.5) -- (-3.5, 3.2) node[above] {$z$};

            % BACK CYLINDER (D6_5 left half, perfectly aligned at z = 0.3)
            \shade[top color=blue!30, bottom color=blue!70] 
                (-3, 0.5) -- (0, 0.5) arc (90:-90:0.1 and 0.2) -- (-3, 0.1) arc (-90:-270:0.1 and 0.2) -- cycle;
            \filldraw[fill=blue!20, draw=blue!50!black, thick] (-3, 0.3) ellipse (0.1 and 0.2);
            \draw[blue!50!black, thick] (-3, 0.5) -- (0, 0.5);
            \draw[blue!50!black, thick] (-3, 0.1) -- (0, 0.1);

            % VERTICAL PLANE (D6_7)
            % Top edge lowered to 2.8 so it doesn't hit the top label
            \filldraw[fill=red!30, draw=red!60!black, thick, opacity=0.75] 
                (-1.5, -2) -- (1.5, -1) -- (1.5, 2.8) -- (-1.5, 1.8) -- cycle;

            % INTERSECTION CIRCLE ON PLANE
            \draw[blue!50!black, dashed, thick] (0, 0.3) ellipse (0.15 and 0.2);

            % FRONT CYLINDER (D6_5 right half, perfectly aligned at z = 0.3)
            \shade[top color=blue!30, bottom color=blue!70, opacity=0.9] 
                (0, 0.5) -- (3, 0.5) arc (90:-90:0.1 and 0.2) -- (0, 0.1) arc (-90:90:0.15 and 0.2) -- cycle;
            \filldraw[fill=blue!20, draw=blue!50!black, thick] (3, 0.3) ellipse (0.1 and 0.2);
            \draw[blue!50!black, thick] (0, 0.5) -- (3, 0.5);
            \draw[blue!50!black, thick] (0, 0.1) -- (3, 0.1);

            % LABELS
            \node[above, red!60!black] at (0, 3.2) {D$6_7$};
            \node[above, blue!80!black] at (2.2, 0.6) {D$6_5$ ($z_0$)};
            \node[above, blue!80!black] at (-2.2, 0.6) {D$6_5$ ($z_0$)};
            \node[below left] at (0, -2.5) {$\alpha = 0$};
        \end{tikzpicture}
    \end{minipage}\hfill
    % ==========================================
    % RIGHT PANEL: Recombined (alpha != 0)
    % ==========================================
    \begin{minipage}{0.48\textwidth}
        \centering
        \begin{tikzpicture}[scale=1]
            % Axes
            \draw[->, thick, gray] (-3.5, -2.5) -- (3.5, -2.5) node[right] {$x_5$};
            \draw[->, thick, gray] (-3.5, -2.5) -- (-3.5, 3.2) node[above] {$z$};

            % LEFT FUNNEL (Back)
            % Straight tube from x=-3 to x=-2.2, then flared upward/downward to plane edge at x=-1.5
            \shade[right color=red!30, left color=blue!50] 
                (-1.5, 0.4) to[out=-90, in=0, looseness=0.9] (-2.2, -0.3) -- (-3, -0.3) 
                arc (90:270:0.1 and 0.2) 
                -- (-2.2, -0.7) to[out=0, in=90, looseness=0.9] (-1.5, -1.4) 
                -- cycle;
            \filldraw[fill=blue!20, draw=blue!50!black, thick] (-3, -0.5) ellipse (0.1 and 0.2);
            \draw[blue!50!black, thick] (-3, -0.3) -- (-2.2, -0.3) to[out=0, in=-90, looseness=0.9] (-1.5, 0.4);
            \draw[blue!50!black, thick] (-3, -0.7) -- (-2.2, -0.7) to[out=0, in=90, looseness=0.9] (-1.5, -1.4);

            % VERTICAL PLANE (Recombined D6)
            \filldraw[fill=red!30, draw=red!60!black, thick] 
                (-1.5, -2) -- (1.5, -1) -- (1.5, 2.8) -- (-1.5, 1.8) -- cycle;

            % RIGHT FUNNEL (Front)
            % Straight tube from x=3 to x=1.5, then flared upward/downward to plane center at x=0
            \shade[left color=red!30, right color=blue!50] 
                (0, 2.1) to[out=-90, in=180, looseness=0.9] (1.5, 1.3) -- (3, 1.3) 
                arc (90:-90:0.1 and 0.2) 
                -- (1.5, 0.9) to[out=180, in=90, looseness=0.9] (0, 0.1) 
                -- cycle;
            \filldraw[fill=blue!20, draw=blue!50!black, thick] (3, 1.1) ellipse (0.1 and 0.2);
            \draw[blue!50!black, thick] (3, 1.3) -- (1.5, 1.3) to[out=180, in=-90, looseness=0.9] (0, 2.1);
            \draw[blue!50!black, thick] (3, 0.9) -- (1.5, 0.9) to[out=180, in=90, looseness=0.9] (0, 0.1);

            % LABELS
            \node[above, red!60!black] at (0, 3.2) {Smoothed D6-brane};
            \node[above, blue!80!black] at (2.2, 1.4) {$z^{(+)}$};
            \node[above, blue!80!black] at (-2.2, -0.2) {$z^{(-)}$};
            \node[below left] at (0, -2.5) {$\alpha \neq 0$};
            
            % Dashed guide to show misalignment
            \draw[dashed, gray] (-3, -0.5) -- (3, -0.5);
            \draw[dashed, gray] (-3, 1.1) -- (3, 1.1);
        \end{tikzpicture}
    \end{minipage}
    
    \caption{\textbf{Left:} The T-dual of a D$5$ is a cylindrical D6 (denoted D6$_5$), which pierces the T-dual of a D$7$ (denoted D6$_7$) plane at a single root $z_0$. \textbf{Right:} The T-dual of the fragmentation of the D5 shows up as two semi-infinite funnels, centered around the roots $z^{(\pm)}$. The full system is now a recombined D6-brane.}
    \label{fig:abelian_recombination}
\end{figure}
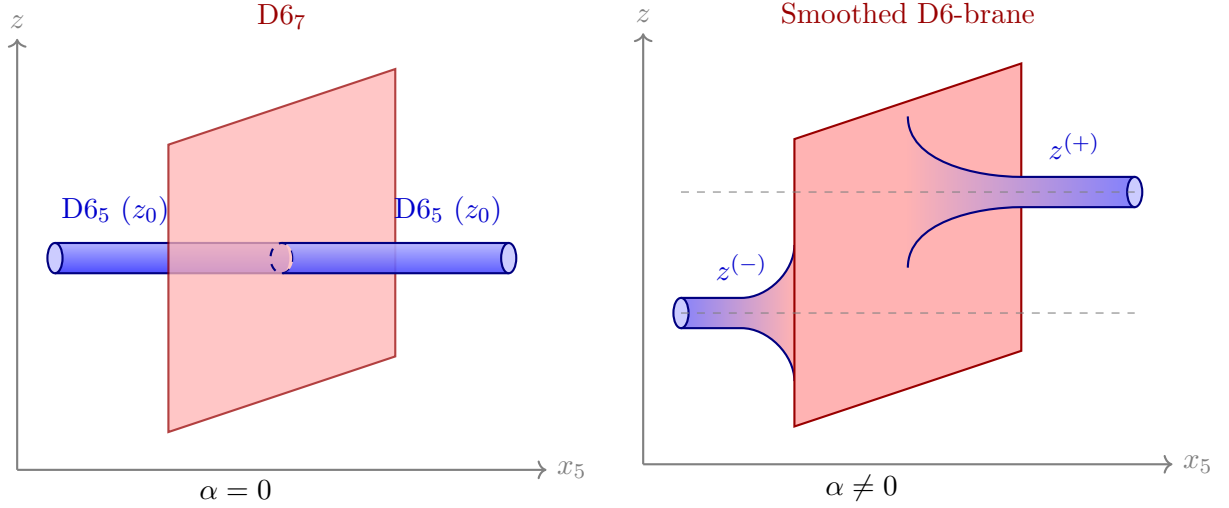

We will now switch to single-valued coordinates by exponentiating. Throughout this section and \S\ref{sec:multiple}--\S\ref{sec:sheaves} there is no NS5-brane, and we work on the dual cylinder
\begin{equation}\label{eq:cyl-frame}
    w_\pm:=e^{\pm (x_5+i \tilde x_9)/\tilde R} \, \quad\text{such that}\quad w_+ w_- = 1\,.
\end{equation}
Then, a D7-brane with $q_L$ and $q_R$ D5-branes shooting off to the left and right, respectively, described by 
\begin{equation} \label{holomlogsolution}
    \Phi = \tilde R \,\sum_{i=1}^{q_L} \log\!\left(\frac{z-z^{(-)}_i}{\Lambda}\right)
           - \tilde R\, \sum_{j=1}^{q_R} \log\!\left(\frac{z-z^{(+)}_j}{\Lambda}\right)
           + \tilde R \, \log c_0\,,
\end{equation}
will be described in the T-dual picture by a single D6-brane wrapping a Riemann surface in $\mathbb{C}^* \times \mathbb{C}$ given by the polynomial hypersurface:
\begin{equation} \label{eq:generaloned7}
    w_+ \left(\prod_{j=1}^{q_R} (z-z^{(+)}_j)\right)= C\,\prod_{i=1}^{q_L} (z-z^{(-)}_i)\,,
\end{equation}
where $C$ encapsulates all the integration constants, crucially including the asymptotic position of the original D7-brane.
As a quick sanity check, notice what happens when we force the D5-branes ending points from the left and the right to coincide, i.e. $z_i^{+} = z_i^{-}$, in the balanced case $q_R = q_L$. The T-dual Riemann surface describing the D6 becomes
\begin{equation}\label{eq:reducible}
      w_+ \left(\prod_{j=1}^{q_R} (z-z^{(+)}_j)\right)= C\,\prod_{i=1}^{q_L} (z-z^{(-)}_i)\, \quad\longrightarrow (w_+-C) \left(\prod_i^q (z-z_i)\right) = 0\,.
\end{equation}
This has a beautiful interpretation: The two semi-infinite D5-branes shooting to the left and right have recombined into connected, infinite D5's that no longer terminate on, but intersect the D7-brane.
In the IIA picture, this translates into having a reducible Riemann surface, as we see in equation \eqref{eq:reducible}. This is despicted in figure \ref{fig:abelian_recombination}.

\paragraph{A note on nomenclature:} In order to streamline the presentation, we will often refer to the T-duals of the D7's and D5's as D$6_7$'s and D$6_5$'s, respectively.

\subsection{M-theory uplift}
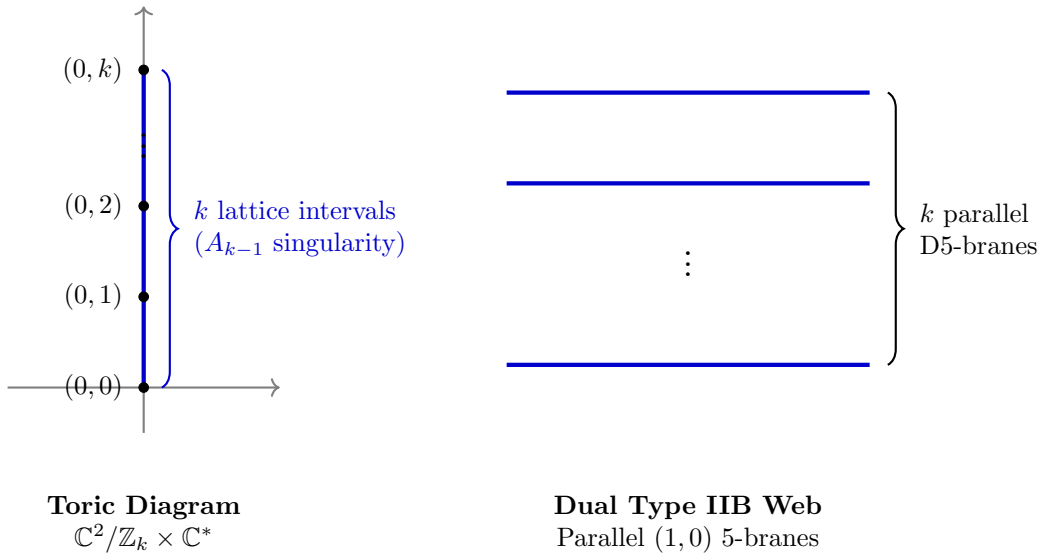
\begin{figure}[h!]
    \centering
    \begin{tikzpicture}[scale=1.2, every node/.style={font=\small}]

        % ==========================================
        % LEFT PANEL: Toric Diagram
        % ==========================================
        \begin{scope}[shift={(0, 0)}]
            % Axes (showing just the vertical domain for the 1D toric graph)
            \draw[->, thick, gray] (-1.5, 0) -- (1.5, 0);
            \draw[->, thick, gray] (0, -0.5) -- (0, 4.2);

            % 1D Toric Graph (A straight vertical line segment)
            \draw[ultra thick, blue!80!black] (0,0) -- (0, 3.5);

            % Lattice Points
            \filldraw[black] (0,0) circle (1.5pt) node[left=4pt] {$(0,0)$};
            \filldraw[black] (0,1) circle (1.5pt) node[left=4pt] {$(0,1)$};
            \filldraw[black] (0,2) circle (1.5pt) node[left=4pt] {$(0,2)$};
            
            % The "dot dot dot" to represent arbitrary k
            \node[font=\Large] at (0, 2.75) {$\vdots$};
            
            \filldraw[black] (0,3.5) circle (1.5pt) node[left=4pt] {$(0,k)$};

            % Multiplicity Brace (on the right side of the axis)
            \draw[decorate, decoration={brace, amplitude=6pt}, thick, blue!80!black] 
                (0.2, 3.5) -- (0.2, 0) node[midway, right=8pt, align=left] {$k$ lattice intervals \\ ($A_{k-1}$ singularity)};
                
            % Subtitle
            \node[align=center] at (0, -1.5) {\textbf{Toric Diagram}\\ $\mathbb{C}^2/\mathbb{Z}_k \times \mathbb{C}^*$};
        \end{scope}

        % ==========================================
        % RIGHT PANEL: Dual 5-Brane Web
        % ==========================================
        \begin{scope}[shift={(6.0, 1.75)}]
            
            % k parallel D5-branes (horizontal)
            \draw[ultra thick, blue!80!black] (-2, 1.5) -- (2, 1.5);
            \draw[ultra thick, blue!80!black] (-2, 0.5) -- (2, 0.5);
            
            % The "dot dot dot" 
            \node[font=\Large] at (0, -0.3) {$\vdots$};
            
            \draw[ultra thick, blue!80!black] (-2, -1.5) -- (2, -1.5);
            
            % Multiplicity Brace
            \draw[decorate, decoration={brace, amplitude=6pt}, thick, black] 
                (2.2, 1.5) -- (2.2, -1.5) node[midway, right=8pt, align=left] {$k$ parallel \\ D5-branes};
            
            % Subtitle
            \node[align=center] at (0, -3.25) {\textbf{Dual Type IIB Web}\\ Parallel $(1,0)$ 5-branes};
        \end{scope}

    \end{tikzpicture}
    \caption{The 1D toric diagram for $\mathbb{C}^2/\mathbb{Z}_k \times \mathbb{C}^*$ (left) and its dual Type IIB 5-brane web (right).}
    \label{fig:toric_Zk_Cstar}
\end{figure}

Having setup the problem purely in terms of IIA on $\mathbb{C}^* \times \mathbb{C}$ witih D6-branes, the M-theory uplift becomes pure geometry. Specifically, it becomes a Taub-NUT fibration over the IIA base, with degeneration locus right over the D6-branes. This can be readily packed into an algebraic variety as follows:
\[
u v= \Delta_{\rm D6}(w_\pm, z)\,,
\]
where $\Delta_{\rm D6}$ is the polynomial covering the Riemann surface wrapped by the D6-brane system. For instance, we could have
\[
uv =z^k\,.
\]
This is the M-theory dual to $k$ D5-branes, and indeed  we recognize this geometry as $\mathbb{C}^2/\mathbb{Z}_k \times \mathbb{C}^*$, which has toric diagram and web diagram depicted in figure \ref{fig:toric_Zk_Cstar}

Now we can consider bringing in some D7-branes from the far $x_5 = \infty$ to a finite distance. Depending on the configuration we want, this will deform the geometry, breaking toricity. One simple example is, take all $k$ D5-branes ending on the same D7. From our previous section, we know that the T-dual D6-curve corresponds to 
\begin{equation} \label{eq:simplejunction}
    \Delta_{\rm D6}: = w_+-\alpha\,z^k\,.
\end{equation}
The M-theory uplift is remarkably simple, it is defined as the algebraic threefold given by
\[
u v = w_+-\alpha\,z^k\,.
\]
In this case, the variety turns out to be non-singular. 

We could have more intricate partitions of the D5-branes as they end on various D7-branes. Generally, we could have something like
\[
u v = \prod_{i=1}^N (w_+-\alpha\,z^{\lambda_i})\,,
\]
with $\sum \lambda_i = k$. Much of this is reminiscent of what was done in \cite{Bourget:2023wlb}, but now we're deriving this from first principles, and do not resort to resolutions. 

In the next section, we will show that there is some structure to these polynomials. Namely, by interpreting them as spectral equations, we can read off symmetry patterns, and understand these phenomena as brane recombination, or \emph{T-brane} states.

\section{Multiple D7-branes}\label{sec:multiple}
Now we will study the T-dual of D5's ending on multiple D7-branes.
\subsection{T-dual of semi-infinite D5's ending on multiple D7-branes}
Let us now generalize to T-dualizing $k$ D5-branes ending on $N$ D7-branes. Let the $k$ 5-branes end on the $N$ 7-branes from the left, according to a partition $[\lambda] = [\lambda_1, \dots, \lambda_N]$, with $\sum_{i=1}^{N} \lambda_i=k$.
We achieve this by essentially constructing a reducible Riemann surface:
\begin{equation}
\Sigma_{\rm red}:= \quad \prod_{a=1}^{N} \left(w_+-C_a \prod_{i=1}^{\lambda_a}(z-z^{(a)}_i) \right)=0\,.
\end{equation}

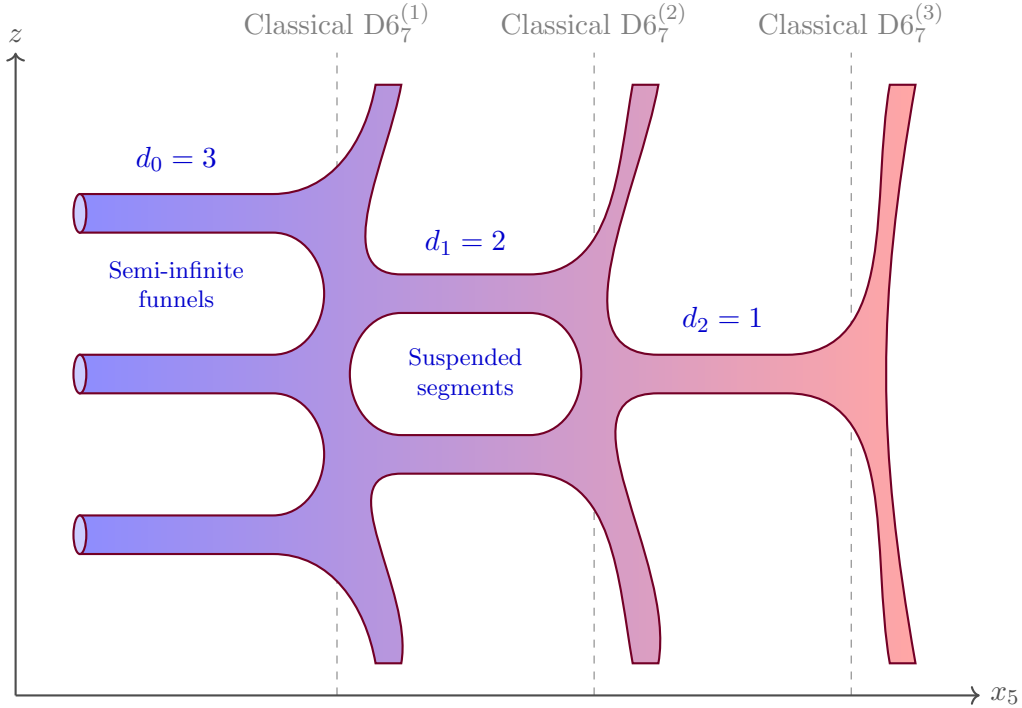
\begin{figure}[h!]
    \centering
    \begin{tikzpicture}[scale=0.85]
        
        % Background Classical Loci
        \draw[dashed, gray] (-1, -5) -- (-1, 5) node[above] {Classical D6$_7^{(1)}$};
        \draw[dashed, gray] (3, -5) -- (3, 5) node[above] {Classical D6$_7^{(2)}$};
        \draw[dashed, gray] (7, -5) -- (7, 5) node[above] {Classical D6$_7^{(3)}$};

        % -----------------------------------------------------
        % DEFINE THE OUTER CONTINUOUS BOUNDARY OF THE BRANE
        % -----------------------------------------------------
        \def\outerpath{
            % Top of T0a
            (-5, 2.8) -- (-2, 2.8)
            % Flare up to Plane 1 Top Left
            to[out=0, in=-100] (-0.4, 4.5)
            % Across Plane 1 Top
            -- (0, 4.5)
            % Flare down to T1a Top Left
            to[out=-100, in=180] (0, 1.55)
            % Across T1a Top
            -- (2, 1.55)
            % Flare up to Plane 2 Top Left
            to[out=0, in=-100] (3.6, 4.5)
            % Across Plane 2 Top
            -- (4.0, 4.5)
            % Flare down to T2a Top Left
            to[out=-100, in=180] (4, 0.3)
            % Across T2a Top
            -- (6, 0.3)
            % Flare up to Plane 3 Top Left
            to[out=0, in=-100] (7.6, 4.5)
            % Across Plane 3 Top
            -- (8.0, 4.5)
            % Plane 3 Right Edge (bulges left in the middle)
            to[out=-100, in=100] (8.0, -4.5)
            % Across Plane 3 Bottom
            -- (7.6, -4.5)
            % Flare up to T2a Bottom Right
            to[out=100, in=0] (6, -0.3)
            % Across T2a Bottom
            -- (4, -0.3)
            % Flare down to Plane 2 Bottom Right
            to[out=180, in=80] (4.0, -4.5)
            % Across Plane 2 Bottom
            -- (3.6, -4.5)
            % Flare up to T1b Bottom Right
            to[out=100, in=0] (2, -1.55)
            % Across T1b Bottom
            -- (0, -1.55)
            % Flare down to Plane 1 Bottom Right
            to[out=180, in=80] (0, -4.5)
            % Across Plane 1 Bottom
            -- (-0.4, -4.5)
            % Flare up to T0c Bottom Right
            to[out=100, in=0] (-2, -2.8)
            % Across T0c Bottom
            -- (-5, -2.8)
            % Up the opening of T0c
            -- (-5, -2.2)
            % Across T0c Top
            -- (-2, -2.2) 
            % Void between T0c and T0b (Plane 1 left edge bulge)
            to[out=0, in=-90] (-1.2, -1.25) to[out=90, in=0] (-2, -0.3)
            % Across T0b Bottom
            -- (-5, -0.3)
            % Up the opening of T0b
            -- (-5, 0.3)
            % Across T0b Top
            -- (-2, 0.3) 
            % Void between T0b and T0a (Plane 1 left edge bulge)
            to[out=0, in=-90] (-1.2, 1.25) to[out=90, in=0] (-2, 2.2)
            % Across T0a Bottom
            -- (-5, 2.2)
            % Up the opening of T0a to close
            -- cycle
        }

        % -----------------------------------------------------
        % DEFINE THE CLOSED HOLE (VOID BETWEEN T1a AND T1b)
        % -----------------------------------------------------
        \def\voidthree{
            % Bottom of T1a
            (0, 0.95) -- (2, 0.95)
            % Right edge of hole (Plane 2 left edge bulge)
            to[out=0, in=90] (2.8, 0) to[out=-90, in=0] (2, -0.95)
            % Top of T1b
            -- (0, -0.95)
            % Left edge of hole (Plane 1 right edge bulge)
            to[out=180, in=-90] (-0.8, 0) to[out=90, in=180] (0, 0.95)
            -- cycle
        }

        % -----------------------------------------------------
        % RENDER THE SURFACE
        % -----------------------------------------------------
        % Fill with 3D gradient and cut out the hole
        \shade[left color=blue!45, right color=red!35, even odd rule] \outerpath \voidthree ;
        
        % Stroke the complex boundary
        \draw[thick, purple!60!black, even odd rule] \outerpath \voidthree ;

        % -----------------------------------------------------
        % ADD 3D PERSPECTIVE ELLIPSES AT THE OPENINGS
        % -----------------------------------------------------
        \filldraw[fill=blue!20, draw=purple!60!black, thick] (-5, 2.5) ellipse (0.1 and 0.3);
        \filldraw[fill=blue!20, draw=purple!60!black, thick] (-5, 0) ellipse (0.1 and 0.3);
        \filldraw[fill=blue!20, draw=purple!60!black, thick] (-5, -2.5) ellipse (0.1 and 0.3);

        % -----------------------------------------------------
        % FOREGROUND AXES & LABELS
        % -----------------------------------------------------
        \draw[->, thick, black!70] (-6, -5) -- (9, -5) node[right] {$x_5$};
        \draw[->, thick, black!70] (-6, -5) -- (-6, 5) node[above] {$z$};

        % Brane degree annotations
        \node[above, blue!80!black, font=\bfseries] at (-3.5, 3.0) {$d_0 = 3$};
        \node[above, blue!80!black, font=\bfseries] at (1.0, 1.7) {$d_1 = 2$};
        \node[above, blue!80!black, font=\bfseries] at (5.0, 0.5) {$d_2 = 1$};
        
        \node[align=center, blue!80!black, font=\footnotesize] at (-3.5, 1.4) {Semi-infinite \\ funnels};
        \node[align=center, blue!80!black, font=\footnotesize] at (1.0, 0) {Suspended \\ segments};

    \end{tikzpicture}
    \caption{The geometric realization of a general recombined D6-brane representing the T-dual of a non-trivial distribution of D5's ending on D7's. Three semi-infinite D$6_5$ funnels ($d_0=3$) enter from the left and merge into the first D$6_7$. The fragments result in two suspended funnels ($d_1=2$) connecting to the second D$6_7$, and finally a single funnel ($d_2=1$) connecting to the third D$6_7$. The entire configuration forms a single, perfectly smooth holomorphic curve. The vertical branes bend logarithmically towards the right ($x_5 \to +\infty$) at large $|z|$.}
    \label{fig:t_brane_partition_recombined}
\end{figure}

Note that this reducible Riemann surface can be deformed to a more generic, reducible one of the form:
\begin{equation}\label{eq:irredriemann} \Sigma_{\rm gen}:= \quad \sum_{a=0}^N w_+^a P_a(z)
\end{equation}
where deg$(P_a(z)) = d_a$ is such that $d_a-d_{a+1} = \lambda_a$, with $d_N=0$, and $d_0=k$. By choosing the $C_a$, we can make each polynomial $P_a$ monic. This brings the total number of moduli to $\sum_a d_a$,
which is exactly the number of brane segments. This matches the standard picture of suspended branes (including the semi-infinite ones in the count). This $\Sigma_{\rm gen}$ is therefore the T-dual to the IIB situation where the branes have fractured along the D7-branes, as shown in figure \ref{fig:t_brane_partition_recombined}

A QFT argument by Witten \cite{Witten:1997sc}, after some translation, shows that the degree in $w_+$ gives the number of D6$_7$'s, and that the degree $d_a$ gives the number of D6$_5$'s in the $[a, a+1]$ `bin', so-to-speak. The degrees $d_0$ and $d_N$ give the number of semi-infinite D6$_5$'s shooting to the left and right, respectively. So
\begin{align} \label{eq:degreerules}
    \deg(w_+) &= \text{\# D6$_7$-branes} \nonumber \\
    d_a &= 
    \text{\# suspended D6$_5$-branes between D6$_7^{a}$ and D6$_7^{a+1}$}
\end{align}
We will recover this fact later on when we introduce the coherent sheaves picture.

\subsection{Suspended branes}
A special case of the previous section is the T-dual of a situation where finite D5-branes are suspended between D7-branes, with no semi-infinite D5's in either direction. It is extremely simple to describe, and the answer is (in a different duality frame) already in \cite{Witten:1997sc}: We simply set $P_0 = 1$. 

Here's how we understand this. Suppose, by contradiction, $P_0(z) \neq$ const, then it has at least one root, say at $z=z_0$. On that divisor, the equation for the Riemann surface restricts to a reducible equation:
\[
\Sigma_{\rm gen} \cap \{z=z_0\}:= \quad w_+ \, \left(\sum_{a=1}^N w_+^{a-1} P_a(z_0) \right) = 0\,.
\]
The branch $w_+ = 0$ means that the recombined object shoots off into $x_5 \rightarrow -\infty$, which is interpreted as a semi-infinite D5-brane shooting off to the left in the dual IIB picture.

\section{Branes as sheaves and T-branes}\label{sec:sheaves}
In order to unify the configurations in the previous section, and take things one step further, we will introduce the language of tachyon condensation, or mathematically, coherent sheaves. This approach proved successful in \cite{Bourget:2023wlb}. Here we will make slightly different use of it, specifically tailored to linking our non-Abelian partitions to the geometric spectral curves, and again, we will not be resorting to blow-ups.

The beauty of this formalism is that it will allow us to construct brane junctions by simply gluing building blocks: We will initially configure the D6$_5$'s and D6$_7$'s as intersecting branes, and subsequently glue them by activating bifundamental strings, in the form of \emph{T-brane} data.
\subsection{Abelian brane recombination}
Let us begin with the simplest possible situation: Take an infinite D5 that intersects a D7, allow the D5 to fragment at the junction. What is the T-dual of that process?

First, define the T-dual of the D7-brane as a coherent sheaf:
\begin{equation}
    0 \longrightarrow \mathcal{O} \xrightarrow{\,T_7:=(w_+-C)\,} \mathcal{O} \longrightarrow \mathcal{O}_{{\rm D}6_7} \longrightarrow 0\,,
\end{equation}
where the first term is an anti-D8, the second a D8, the map is the \emph{tachyon field}, and the third term is the cokernel sheaf, representing the remnant of the tachyon condensation process. We refer to \cite{Sharpe:2003dr} for a complete introduction, and to \cite{Collinucci:2014qfa} for a more concise and targeted one. This coherent sheaf represents the D6-brane, call it ${\rm D}6_7$ at the locus
\[
{\rm D6}_7: \quad \{w_+-C = 0\}\,,
\]
Now consider the T-dual of the infinite D5-brane to be the ${\rm D}6_5$ at 
\[
{\rm D6}_5: \quad \{z-z_0 = 0\}\,.
\]
This is represented as the following coherent sheaf
\begin{equation}
    0 \longrightarrow \mathcal{O} \xrightarrow{\,T_5:=(z-z_0)\,} \mathcal{O} \longrightarrow \mathcal{O}_{{\rm D}6_5} \longrightarrow 0\,,
\end{equation}
The subscripts '7' and '5' are there to keep track of the IIB duals of each D6-brane. Now we combine the system into one coherent sheaf:
\begin{equation}
    0 \longrightarrow \mathcal{O}^{\oplus 2} \xrightarrow{\,\mathcal{T}\,} \mathcal{O}^{\oplus 2} \longrightarrow \mathcal{O}_{{\rm D}6_5\,\cup\, {\rm D}6_7} \longrightarrow 0\,,
\end{equation}
with $\mathcal{T} = T_5 \oplus T_7$
\[
\mathcal{T} = \begin{pmatrix}
    z-z_0 & 0 \\ 0 & w_+-C
\end{pmatrix}\,.
\]
Evaluating the determinant of this matrix gives the expected reducible curve \[\det(\mathcal{T}) = (z-z_0) \, (w_+-C)\,.\]
Now consider switching on one bifundamental D$6_5$/D$6_7$-string. Those correspond to offdiagonal fluctuations we can add to $\mathcal{T}$, spoiling the direct sum structure:
\[ \label{eq:recombintersecting}
\mathcal{T}_{\rm recomb} = \begin{pmatrix}
    z-z_0 & 1 \\ \alpha & w_+-C
\end{pmatrix}\,,
\]
which has irreducible determinant of the form
\begin{align}
    \det(\mathcal{T}_{\rm recomb}) &= w_+ (z-z_0)+(C (z-z_0)-\alpha))\,.
\end{align}
This matches the logic of \cite{Witten:1997sc}: It has the form
\[w_+ P_1(z)+P_0(z)\,,\] where we recognize one D6$_7$ (degree one in $w_+$), with one D6$_5$ shooting off to the right (deg$(P_1(z)=1$), and one to the left (deg$(P_0(z)=1$).
The T-brane data in question, in the off-diagonal entries, recombine the intersecting branes, in such a way that the left and right-shooting semi-infinite branes do not coincide. The result is depicted in \ref{fig:abelian_recombination}

\subsection{Brane ending on brane}
\begin{figure}[h!]
    \centering
    \begin{tikzpicture}[scale=1]
        % Axes
        \draw[->, thick, gray] (-4.5, -3.5) -- (3.5, -3.5) node[right] {$x_5$};
        \draw[->, thick, gray] (-4.5, -3.5) -- (-4.5, 3.5) node[above] {$z$};

        % The Trumpet Surface (Elongated bent brane)
        % Drawn as a single continuous surface smoothly expanding from a tube into a plane
        \shade[left color=blue!50, right color=red!30]
            (-4, 0.2) -- (-1.5, 0.2) .. controls (0, 0.2) and (1, 1) .. (2.5, 3)
            arc (90:-90:0.4 and 3)
            .. controls (1, -1) and (0, -0.2) .. (-1.5, -0.2) -- (-4, -0.2)
            arc (-90:-270:0.1 and 0.2);
            
        % Outlines to emphasize the logarithmic flare
        \draw[thick, purple!60!black] (-4, 0.2) -- (-1.5, 0.2) .. controls (0, 0.2) and (1, 1) .. (2.5, 3);
        \draw[thick, purple!60!black] (-4, -0.2) -- (-1.5, -0.2) .. controls (0, -0.2) and (1, -1) .. (2.5, -3);
        
        % Left opening (Semi-infinite D6_5 tube coming from the left)
        \filldraw[fill=blue!20, draw=blue!50!black, thick] (-4, 0) ellipse (0.1 and 0.2);
        
        % Right opening lip (front of the flared D6_7)
        \draw[thick, red!60!black] (2.5, 3) arc (90:-90:0.4 and 3);
        % Right opening lip (back, dashed to show 3D volume)
        \draw[thick, dashed, red!60!black] (2.5, 3) arc (90:270:0.4 and 3);

        % Labels and Annotations
        \node[above, blue!80!black] at (-2.5, 0.3) {Semi-infinite D$6_5$ tube};
        \node[left, red!60!black] at (1.5, 3) {Logarithmic flare};
        \node[right, red!60!black, align=left] at (3.0, 0) {Asymptotic \\ D$6_7$ plane};
        
        % Math annotation pointing to the flare
        \draw[->, thick, red!60!black] (0.5, 2) to[out=180, in=90] (0.2, 0.6);
        \node[above, red!60!black] at (0.5, 2) {$x_5 \sim R \log|z|$};
        
        % Dashed center line to show the classical D7 position
        \draw[dashed, gray] (0, -3) -- (0, 3);
        \node[below, gray] at (0, -3) {Classical D$6_7$ locus};

    \end{tikzpicture}
    \caption{The T-dual of a semi-infinite D5 coming from $x_5 = -\infty$, ending on a D7: A single, smooth, elongated D6-brane captures the entire physics of the junction.}
    \label{fig:elongated_bent_brane}
\end{figure}
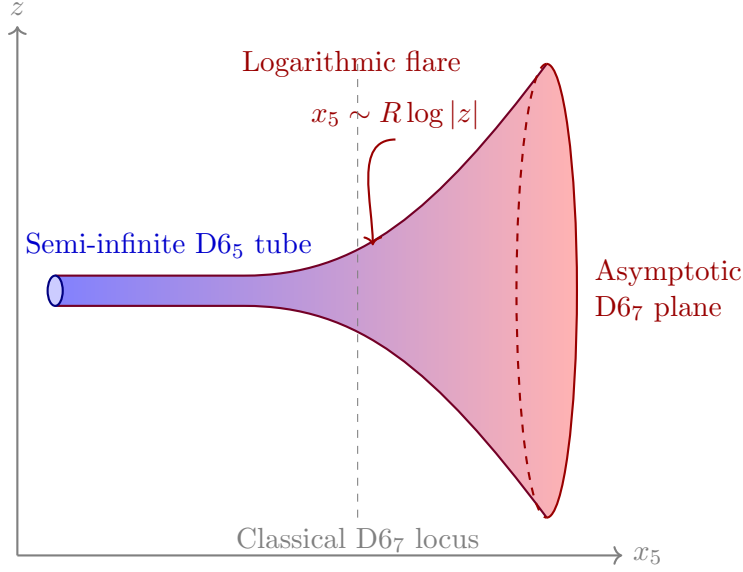

Let us now engineer the T-dual of a semi-infinite D5 ending on a D7 from the left, as shown in figure \ref{fig:elongated_bent_brane}.
We start with a similar D$6_5$/D$6_7$ intersecting Ansatz as \ref{eq:recombintersecting}, but we move the D6$_7$ infinitely to the right, i.e. $x_5 \rightarrow +\infty$. Intuitively, we do this so that the recombined system has no 'room' for right-most portion of the D$6_5$ brane that shoots off to the right. Concretely, instead of placing it at $w_+-C=0$, we place it at $w_-=0$, which corresponds to $x_5 \rightarrow \infty$. The tachyon now has the form
\[ \label{eq:recombending}
\mathcal{T}_{\rm recomb} = \begin{pmatrix}
    z-z_0 & 1 \\ \rule{0pt}{3ex} C^{-1} & w_-
\end{pmatrix}\,.
\]
The spectral equation for this is 
\[
w_-\,(z-z_0)-C^{-1} = 0 \qquad \cong \qquad w_+-C\,(z-z_0) = 0
\]
where the equivalence is obtained by multiplying the whole equation by $w_+$, which won't change the zero-locus since $w_+ \in \mathbb{C}^*$.
This equation reproduces \ref{eq:generaloned7} for the case $q_R=0, q_L=1$, with $\alpha = C^{-1}.$

\subsection{General non-Abelian recombination warmup: Two branes}
\paragraph{Factorizable case:} Let us consider two D6$_5$'s ending on two D6$_7$'s, and then transition to the case where the two \df's end on a single \ds.

To accomplish this, we essentially take the direct sum of two copies of \eqref{eq:recombending}:
\[
\mathcal{T}_{k=2, N=2} = \left( \begin{array}{cc|cc} 
    z-z_1&0&1&0\\
    0&z-z_2&0&1\\ \hline \rule{0pt}{3ex}
    C_1^{-1}&0&w_-&0\\
    0&C_2^{-1}&0&w_-
\end{array} \right)
\]
which has spectral equation (up to multiplication by $w_+^2$) 
\[C_1\,C_2\, w_+^2\,\det(\mathcal{T}_{k=2, N=2}) = w_+^2 -w_+ \left(C_1 (z-z_1)+C_2 (z-z_2) \right)+\left(C_1
\, C_2\,(z-z_1)\,(z-z_2)\right)\,.\]
The zero locus of this spectral equation describes exactly what we expect: Two 'bins' defined by two \ds-branes, with one \df suspended between them, and two semi-infinite \df's shooting off to the left. This obeys the rules we established in \ref{eq:degreerules}, with $(d_2, d_1, d_0) = (0,1,2)$.

One caveat here: This configuration isn't the most general in its class. The branes are locked in step in the sense that we do not see the separate moduli that allow us to, say, move the suspended brane along the \ds. We will now find the remedy to this.

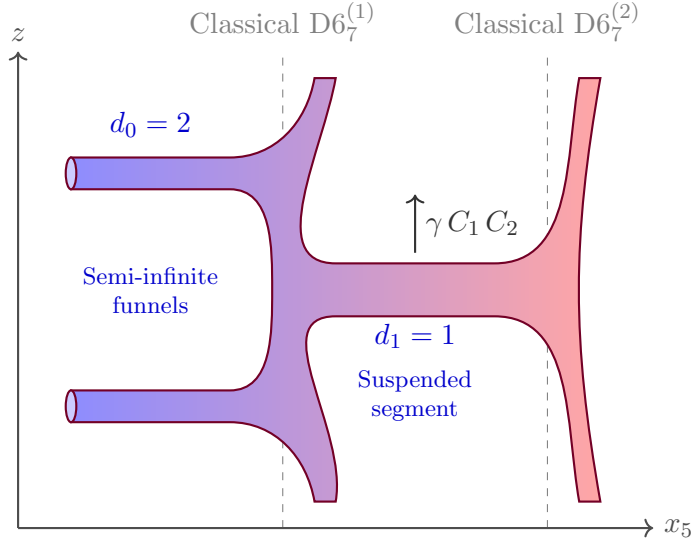
\begin{figure}
    \centering
    \begin{tikzpicture}[scale=0.7]
        
        % Background Classical Loci (Two D6_7 branes)
        \draw[dashed, gray] (-1, -4.5) -- (-1, 4.5) node[above] {Classical D6$_7^{(1)}$};
        \draw[dashed, gray] (4, -4.5) -- (4, 4.5) node[above] {Classical D6$_7^{(2)}$};

        % -----------------------------------------------------
        % DEFINE THE OUTER CONTINUOUS BOUNDARY OF THE BRANE
        % -----------------------------------------------------
        \def\outerpath{
            % Top of T0a (Top semi-infinite funnel)
            (-5, 2.5) -- (-2, 2.5)
            % Flare up to Plane 1 Top Left
            to[out=0, in=-100] (-0.4, 4.0)
            % Across Plane 1 Top
            -- (0, 4.0)
            % Flare down to T1a Top Left (Suspended funnel)
            to[out=-100, in=180] (0, 0.5)
            % Across T1a Top
            -- (3, 0.5)
            % Flare up to Plane 2 Top Left
            to[out=0, in=-100] (4.6, 4.0)
            % Across Plane 2 Top
            -- (5.0, 4.0)
            % Plane 2 Right Edge (bulges left in the middle)
            to[out=-100, in=100] (5.0, -4.0)
            % Across Plane 2 Bottom
            -- (4.6, -4.0)
            % Flare up to T1a Bottom Right
            to[out=100, in=0] (3, -0.5)
            % Across T1a Bottom
            -- (0, -0.5)
            % Flare down to Plane 1 Bottom Right
            to[out=180, in=80] (0, -4.0)
            % Across Plane 1 Bottom
            -- (-0.4, -4.0)
            % Flare up to T0b Bottom Right (Bottom semi-infinite funnel)
            to[out=100, in=0] (-2, -2.5)
            % Across T0b Bottom
            -- (-5, -2.5)
            % Up the opening of T0b
            -- (-5, -1.9)
            % Across T0b Top
            -- (-2, -1.9) 
            % Void between T0b and T0a (Plane 1 left edge bulge)
            to[out=0, in=-90] (-1.2, 0) to[out=90, in=0] (-2, 1.9)
            % Across T0a Bottom
            -- (-5, 1.9)
            % Up the opening of T0a to close
            -- cycle
        }

        % -----------------------------------------------------
        % RENDER THE SURFACE
        % -----------------------------------------------------
        % Fill with 3D gradient
        \shade[left color=blue!45, right color=red!35, even odd rule] \outerpath;
        
        % Stroke the complex boundary
        \draw[thick, purple!60!black, even odd rule] \outerpath;

        % -----------------------------------------------------
        % ADD 3D PERSPECTIVE ELLIPSES AT THE OPENINGS
        % -----------------------------------------------------
        \filldraw[fill=blue!20, draw=purple!60!black, thick] (-5, 2.2) ellipse (0.1 and 0.3);
        \filldraw[fill=blue!20, draw=purple!60!black, thick] (-5, -2.2) ellipse (0.1 and 0.3);

        % -----------------------------------------------------
        % MODULUS ARROW (gamma)
        % -----------------------------------------------------
        % Arrow pointing upwards on the suspended segment
        \draw[->, thick, black!80] (1.5, 0.7) -- (1.5, 1.8) node[midway, right] {$\gamma\,C_1\,C_2$};

        % -----------------------------------------------------
        % FOREGROUND AXES & LABELS
        % -----------------------------------------------------
        \draw[->, thick, black!70] (-6, -4.5) -- (6, -4.5) node[right] {$x_5$};
        \draw[->, thick, black!70] (-6, -4.5) -- (-6, 4.5) node[above] {$z$};

        % Brane degree annotations
        \node[above, blue!80!black, font=\bfseries] at (-3.5, 2.7) {$d_0 = 2$};
        \node[above, blue!80!black, font=\bfseries] at (1.5, -1.3) {$d_1 = 1$};
        
        \node[align=center, blue!80!black, font=\footnotesize] at (-3.5, 0) {Semi-infinite \\ funnels};
        \node[align=center, blue!80!black, font=\footnotesize] at (1.5, -2.0) {Suspended \\ segment};

    \end{tikzpicture}
    \caption{The T-dual of the $[1,1]$-partition after brane fragmentation. The modulus $\gamma$ controls the transverse $z$-position of the suspended segment.}
    \label{fig:t_brane_2_1_partition}
\end{figure}

\paragraph{Recombination:} In order to find the T-dual of the brane fragmentation in IIB, whereby suspended D5-branes start moving `up and down' between two D7-branes independently of the neighboring D5's, (in a way that generates masses for the stretched open strings), we must switch on T-brane data.
\begin{equation}
    \mathcal{T}_{k=2, N=2} = \left( \begin{array}{cc|cc} 
        z-z_1 & 1 & 1 & 0 \\ 
        0 & z-z_2 & 0 & 1 \\ 
        \hline \rule{0pt}{3ex}
        C_1^{-1} & 0 & w_- & 0 \\ 
        \gamma & C_2^{-1} & 0 & w_- 
    \end{array} \right)
\end{equation}
This manifestly breaks the direct sum structure, and the spectral surface no longer factorizes:
\begin{equation} \label{eq:movingsegment}
C_1\, C_2 \,w_+^2\,\det(\mathcal{T}_{k=2, N=2}) = w_+^2 -w_+ \big(C_1 (z-z_1)+C_2 (z-z_2)-\gamma\,C_1\,C_2 \big)+\left(C_1
\, C_2\,(z-z_1)\,(z-z_2)\right)\,.
\end{equation}
Now we see exactly that extra modulus, $\gamma\,C_1\,C_2$ in the $w_+$-linear term, which precisely encodes the movement of the suspended brane. This is illustrated in figure \ref{fig:t_brane_2_1_partition}

\paragraph{Kraft-Procesi transition:}
We can now implement a so-called Kraft-Procesi transition, as understood by \cite{Cabrera:2017njm}, by decoupling the suspended D6$_5$-brane. To accomplish this, we simply send $C_i \mapsto \infty$. The tachyon matrix becomes
\begin{equation} \label{eq:2partition}
    \mathcal{T}_{k=2, N=1} = \left( \begin{array}{cc|cc} 
        z-z_1 & 1 & 1 & 0 \\ 
        0 & z-z_2 & 0 & 1 \\ 
        \hline \rule{0pt}{3ex}
        0 & 0 & w_- & 0 \\ 
        \gamma & 0 & 0 & w_- 
    \end{array} \right)
\end{equation}
with spectral equation
\[(-w_+^2/\gamma)\,\det(\mathcal{T}_{k=2, N=1}) = w_+-\gamma^{-1}\,(z-z_1)\,(z-z_2)\,.\]
Notice how this perfectly reproduces the formula \eqref{eq:generaloned7} for the case $(q_L = 2, q_R=0)$. Notice also, that the limit we took would not make sense if performed directly on the spectral equation \eqref{eq:movingsegment}, but makes perfect sense at the level of the tachyon field.

This transition 
\[ \begin{pmatrix} C_1^{-1} & 0\\\gamma & C_2^{-1}\end{pmatrix} \mapsto \begin{pmatrix} 0 & 0\\\gamma & 0\end{pmatrix}\]
is a transition from a semi-simple matrix to a nilpotent one. More precisely, it corresponds to transitioning from the trivial $[1,1]$ to the $[2]$ partition.

The Schur complements of ${\mathcal T}$ provide a nice visualization what is happening physically:
\begin{equation} \label{eq:2partitionschurup}
    \det(\mathcal{T}) = w_-^2 \det \left[ \begin{pmatrix} z-z_1 & 1 \\ 0 & z-z_2 \end{pmatrix} - \frac{1}{w_-} \begin{pmatrix} 0 & 0 \\ \gamma & 0 \end{pmatrix} \right]\,.
\end{equation}
From the viewpoint of the two \df, the presence of the \ds appears in the form of a \emph{Hitchin pole} on its worldvolume.

Alternatively, we can consider things from the viewpoint of the \ds and write things as follows:
\begin{equation}
    \mathcal{T} = \begin{pmatrix} z\mathbf{1}_2 - \varphi & \tilde{Q} \\ Q & w_- \mathbf{1}_2 \end{pmatrix}
\end{equation}

\begin{equation}
    \det(\mathcal{T}) = \det(z\mathbf{1}_2 - \varphi) \det\left( w_- \mathbf{1}_2 - Q (z\mathbf{1}_2 - \varphi)^{-1} \tilde{Q} \right) = 0\,,
\end{equation}
where, here 
\[\varphi = \begin{pmatrix}
    z-z_1&1\\0&z-z_2
\end{pmatrix}\,.\]
So, the \df's appear as poles on the \ds-worldvolume, and the determinant involves computing the resolvent of the \df-Higgs field. This is reminiscent of the Gaiotto-Witten Nahm poles in 3d setups \cite{Gaiotto:2008ak, Gaiotto:2008sa}, where D3-branes are distributed among various D5-branes according to partitions, which are encoded in boundary conditions for $\mathcal{N}=4, d=4$ SYM.

\subsection{General non-Abelian recombination: Kraft-Procesi transitions}

To capture the generalized Kraft-Procesi transition for an arbitrary partition $[\lambda]$ of $k$ \df's, we map the entire nilpotent hierarchy by starting from the top. We initially align $k$ \df's intersecting $N=k$ \ds's, corresponding to the trivial partition $[1^k]$ with a fully broken $SU(N)$ global symmetry. 

The master tachyon map is a $2k \times 2k$ block matrix where the Higgs field $\varphi$ is diagonal, and the $w_-$ background represents the $k$ distinct \ds spectator branes.
\begin{equation}
    \mathcal{T}_{[1^k]} = \begin{pmatrix} z\mathbf{1}_k - \varphi_{\text{diag}} & \tilde{Q}_{[1^k]} \\ Q_{[1^k]} & w_- \mathbf{1}_k \end{pmatrix}
\end{equation}
with 
\[Q_{[1^k]} = {\rm diag}(C_1^{-1}, \ldots, C_k^{-1})\,, \quad \tilde Q_{[1^k]} = \mathbf{1}_k\,.\]
In order to walk down the Hasse diagram to a coarser partition $[\lambda]$, $\varphi$ deforms into a block-diagonal nilpotent matrix, where each Jordan block has size $\lambda_a$. Crucially, the off-diagonal matrices $Q$ and $\tilde{Q}$ are not arbitrary. 
For a given Jordan block of size $\lambda_a$, the corresponding row vector $Q$ takes the form $(q, 0, \dots, 0)$, while the column vector $\tilde{Q}$ takes $(0, \dots, 0, \tilde{q})^\top$. The irreducible spectral curve is extracted entirely by evaluating the Schur complement over the flavor space:
\begin{equation}
    \det(\mathcal{T}) = \det(z\mathbf{1}_k - \varphi_{[\lambda]}) \det\left( w_- \mathbf{1}_k - Q (z\mathbf{1}_k - \varphi_{[\lambda]})^{-1} \tilde{Q} \right) = 0
\end{equation}
The underlying mechanism generating the geometry lies in the structure of the resolvent matrix $(z\mathbf{1}_k - \varphi_{[\lambda]})^{-1}$. Because $\varphi_{[\lambda]}$ is a Jordan block, inverting $(z\mathbf{1}_k - \varphi_{[\lambda]})$ yields an upper-triangular matrix where the entries gain higher inverse powers of $z$ as one moves away from the diagonal. Specifically, the top-right corner entry develops a maximal pole of order $z^{-\lambda_a}$. 

Because the coupling vectors $Q$ and $\tilde{Q}$ possess non-zero entries only at their respective extremities, they act as algebraic projectors that isolate exactly this top-right corner. Consequently, the matrix multiplication collapses to a single term:
\begin{equation}
    Q (z\mathbf{1}_k - \varphi_{[\lambda]})^{-1} \tilde{Q} = \frac{q\tilde{q}}{z^{\lambda_a}}
\end{equation}
Setting the Schur complement to zero thus yields $w_- - q\tilde{q}/z^{\lambda_a} = 0$. After multiplying by the determinant of the other block, we recover the expected 
$z^{\lambda_a}\,w_- - q\tilde{q} = 0$, which is equivalent to $z^{\lambda_a} - w_+\,q\tilde{q} = 0$, reproducing the expected result from \eqref{eq:generaloned7}.

This mechanism works for general partitions. We simply reduce every system to direct sums of block.
For a partition $\lambda =[\lambda_1, \ldots, \lambda_m$, the $\lambda_i$ block has the following Schur complement form:
\begin{equation}
\mathcal{T}{(\lambda_i)} = \begin{pmatrix} 
    z & -1 & 0 & \dots & 0 \\
    0 & z & -1 & \dots & 0 \\
    \vdots & \vdots & \ddots & \ddots & \vdots \\
    0 & 0 & \dots & z & -1 \\
    -w_+ q \tilde{q} & 0 & \dots & 0 & z 
\end{pmatrix}
\end{equation}
The determinant of this block is cleanly $z^{\lambda_i} - w_+ q \tilde{q}$, perfectly matching the local spectral curve of a $d$-th order branch point. 

For a generic gauge partition $\lambda = [d_1, d_2, \dots, d_m]$, the total matrix is simply the direct sum of these $m$ independent building blocks:
\begin{equation}
\mathcal{T}^{\text{decoupled}} = \bigoplus_{i=1}^m \mathcal{T}^{(\lambda_i)}
\end{equation}
This will give rise to a spectral curve of the form
\[
\det(\mathcal{T})=\prod_{i=1}^m \left(z^{\lambda_i}-w_+\right)\,.
\]
Now we are free to couple these various blocks to each other by switching on further entries in the $Q, \tilde Q$ matrices. This will allow suspended branes to move 'up and down', leading to an irreducible Riemann surface of the form \eqref{eq:irredriemann}.
\paragraph{M-theory uplift}
The M-theory uplift follows a simple recipe:
\[
u v = \det(\mathcal{T})\,.
\]
The $C^* \cong \mathbb{R} \times S^1$ contains a circle, the M-theory circle, that collapses precisely where the tachyon field fails to be invertible, i.e. wherever $\det(\mathcal{T})=0$. 

\subsection{Full example}
\paragraph{Toric case} To conclude this section, we give a concrete example of a web diagram and its non-toric deformation.
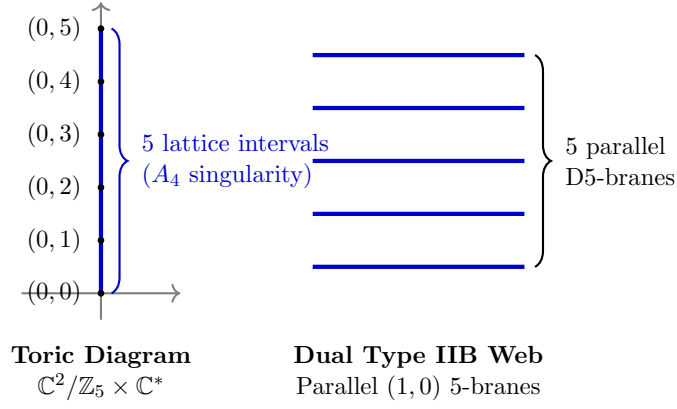
\begin{figure}[h!]
    \centering
    \begin{tikzpicture}[scale=0.7, every node/.style={scale=0.85}]

        % ==========================================
        % LEFT PANEL: Toric Diagram (k=5)
        % ==========================================
        \begin{scope}[shift={(0, 0)}]
            % Axes
            \draw[->, thick, gray] (-1.5, 0) -- (1.5, 0);
            \draw[->, thick, gray] (0, -0.5) -- (0, 5.5);

            % 1D Toric Graph (A straight vertical line segment)
            \draw[ultra thick, blue!80!black] (0,0) -- (0, 5);

            % Explicit Lattice Points for k=5
            \foreach \y in {0,1,2,3,4,5} {
                \filldraw[black] (0,\y) circle (1.5pt) node[left=4pt] {$(0,\y)$};
            }

            % Multiplicity Brace
            \draw[decorate, decoration={brace, amplitude=6pt}, thick, blue!80!black] 
                (0.2, 5) -- (0.2, 0) node[midway, right=8pt, align=left] {5 lattice intervals \\ ($A_4$ singularity)};
                
            % Subtitle
            \node[align=center] at (0, -1.5) {\textbf{Toric Diagram}\\ $\mathbb{C}^2/\mathbb{Z}_5 \times \mathbb{C}^*$};
        \end{scope}

        % ==========================================
        % RIGHT PANEL: Dual 5-Brane Web (k=5)
        % ==========================================
        % Shifted up to y=2.5 to perfectly center it with the left panel
        \begin{scope}[shift={(6.0, 2.5)}]
            
            % 5 explicit parallel D5-branes (horizontal)
            \foreach \y in {2, 1, 0, -1, -2} {
                \draw[ultra thick, blue!80!black] (-2, \y) -- (2, \y);
            }
            
            % Multiplicity Brace
            \draw[decorate, decoration={brace, amplitude=6pt}, thick, black] 
                (2.2, 2) -- (2.2, -2) node[midway, right=8pt, align=left] {5 parallel \\ D5-branes};
            
            % Subtitle (Shifted down to perfectly match the baseline of the left subtitle)
            \node[align=center] at (0, -4.0) {\textbf{Dual Type IIB Web}\\ Parallel $(1,0)$ 5-branes};
        \end{scope}

    \end{tikzpicture}
    \caption{The 1D toric diagram for $\mathbb{C}^2/\mathbb{Z}_5 \times \mathbb{C}^*$ (left) and its dual Type IIB 5-brane web (right), explicitly drawn for $k=5$.}
    \label{fig:toric_Z5_Cstar}
\end{figure}
Figure \ref{fig:toric_Z5_Cstar} corresponds to the purely toric situation: 5 parallel D5-branes, no NS5's, and no D7's. The tachyon will be simply
\[
\mathcal{T} = {\rm diag}(z-z_1, \ldots, z-z_5)\,,
\]
and the M-theory geometry
\[
u v = \prod_{i=1}^5 (z-z_i)\,.\]
Upon pressing the branes together $z_i \mapsto 0$, we get the $\mathbb{C}^2/\mathbb{Z}_5 \times \mathbb{C}^*$ geometry, which is toric.

\begin{figure}[h!]
    \centering
    \scalebox{0.62}{
    \begin{tikzpicture}[scale=0.85]

    % =======================================================
    % PARAMETERS (for reference)
    %   d0 = 5 semi-infinite funnels  (left of Plane 1)
    %   d1 = 4 suspended funnels      (between Plane 1 and Plane 2)
    %   d2 = 3 suspended funnels      (between Plane 2 and Plane 3)
    %   th = 0.3  (half-thickness of every tube)
    %   x1=-1, x2=4.2, x3=9.4  (classical loci), fw = 1.6 (flare width)
    % =======================================================

    % Background Classical Loci
    \draw[dashed, gray] (-1, -6) -- (-1, 6) node[above] {Classical D6$_7^{(1)}$};
    \draw[dashed, gray] (4.2, -6) -- (4.2, 6) node[above] {Classical D6$_7^{(2)}$};
    \draw[dashed, gray] (9.4, -6) -- (9.4, 6) node[above] {Classical D6$_7^{(3)}$};

    % -----------------------------------------------------
    % OUTER CONTINUOUS BOUNDARY OF THE BRANE
    % -----------------------------------------------------
    \def\outerpath{
        % ---- TOP SWEEP (left to right) ----
        % Top of T0_1 (topmost of the 5 semi-infinite funnels)
        (-6, 3.1) -- (-2.6, 3.1)
        % Flare up to Plane 1 top
        to[out=0, in=-100] (-1.4, 5.0)
        -- (-0.6, 5.0)
        % Flare down to top of T1_1 (topmost of the 4 suspended funnels)
        to[out=-100, in=180] (0.6, 2.25)
        -- (2.6, 2.25)
        % Flare up to Plane 2 top
        to[out=0, in=-100] (3.8, 5.0)
        -- (4.6, 5.0)
        % Flare down to top of T2_1 (topmost of the 3 suspended funnels)
        to[out=-100, in=180] (5.8, 1.7)
        -- (7.8, 1.7)
        % Flare up to Plane 3 top (terminal)
        to[out=0, in=-100] (9.0, 5.0)
        -- (9.8, 5.0)
        -- (11.0, 5.0)
        % ---- RIGHT CLOSING CAP (bulges left) ----
        to[out=-100, in=100] (11.0, -5.0)
        % ---- BOTTOM SWEEP (right to left) ----
        -- (9.8, -5.0)
        % Flare to bottom of T2_3 (bottommost of the 3 suspended funnels)
        to[out=100, in=0] (7.8, -1.7)
        -- (5.8, -1.7)
        % Flare to Plane 2 bottom
        to[out=180, in=80] (4.6, -5.0)
        -- (3.8, -5.0)
        % Flare to bottom of T1_4 (bottommost of the 4 suspended funnels)
        to[out=100, in=0] (2.6, -2.25)
        -- (0.6, -2.25)
        % Flare to Plane 1 bottom
        to[out=180, in=80] (-0.6, -5.0)
        -- (-1.4, -5.0)
        % Flare to bottom of T0_5 (bottommost of the 5 semi-infinite funnels)
        to[out=100, in=0] (-2.6, -3.1)
        -- (-6, -3.1)
        % ---- LEFT WEAVE (bottom to top), through the 5 openings ----
        % Up the opening of T0_5
        -- (-6, -2.5)
        % Across T0_5 top
        -- (-2.0, -2.5)
        % Void between T0_5 and T0_4
        to[out=0, in=-90] (-1.2, -2.1) to[out=90, in=0] (-2.0, -1.7)
        % Across T0_4 bottom
        -- (-6, -1.7)
        % Up the opening of T0_4
        -- (-6, -1.1)
        % Across T0_4 top
        -- (-2.0, -1.1)
        % Void between T0_4 and T0_3
        to[out=0, in=-90] (-1.2, -0.7) to[out=90, in=0] (-2.0, -0.3)
        % Across T0_3 bottom
        -- (-6, -0.3)
        % Up the opening of T0_3
        -- (-6, 0.3)
        % Across T0_3 top
        -- (-2.0, 0.3)
        % Void between T0_3 and T0_2
        to[out=0, in=-90] (-1.2, 0.7) to[out=90, in=0] (-2.0, 1.1)
        % Across T0_2 bottom
        -- (-6, 1.1)
        % Up the opening of T0_2
        -- (-6, 1.7)
        % Across T0_2 top
        -- (-2.0, 1.7)
        % Void between T0_2 and T0_1
        to[out=0, in=-90] (-1.2, 2.1) to[out=90, in=0] (-2.0, 2.5)
        % Across T0_1 bottom
        -- (-6, 2.5)
        % Up the opening of T0_1 to close
        -- cycle
    }

    % -----------------------------------------------------
    % ENCLOSED HOLES BETWEEN THE 4 SUSPENDED FUNNELS (T1 GROUP)
    % -----------------------------------------------------
    \def\voidOneA{
        (0.6, 1.65) -- (2.6, 1.65)
        to[out=0, in=90] (3.4, 1.3) to[out=-90, in=0] (2.6, 0.95)
        -- (0.6, 0.95)
        to[out=180, in=-90] (-0.2, 1.3) to[out=90, in=180] (0.6, 1.65)
        -- cycle
    }
    \def\voidOneB{
        (0.6, 0.35) -- (2.6, 0.35)
        to[out=0, in=90] (3.4, 0) to[out=-90, in=0] (2.6, -0.35)
        -- (0.6, -0.35)
        to[out=180, in=-90] (-0.2, 0) to[out=90, in=180] (0.6, 0.35)
        -- cycle
    }
    \def\voidOneC{
        (0.6, -0.95) -- (2.6, -0.95)
        to[out=0, in=90] (3.4, -1.3) to[out=-90, in=0] (2.6, -1.65)
        -- (0.6, -1.65)
        to[out=180, in=-90] (-0.2, -1.3) to[out=90, in=180] (0.6, -0.95)
        -- cycle
    }

    % -----------------------------------------------------
    % ENCLOSED HOLES BETWEEN THE 3 SUSPENDED FUNNELS (T2 GROUP)
    % -----------------------------------------------------
    \def\voidTwoA{
        (5.8, 1.1) -- (7.8, 1.1)
        to[out=0, in=90] (8.6, 0.7) to[out=-90, in=0] (7.8, 0.3)
        -- (5.8, 0.3)
        to[out=180, in=-90] (5.0, 0.7) to[out=90, in=180] (5.8, 1.1)
        -- cycle
    }
    \def\voidTwoB{
        (5.8, -0.3) -- (7.8, -0.3)
        to[out=0, in=90] (8.6, -0.7) to[out=-90, in=0] (7.8, -1.1)
        -- (5.8, -1.1)
        to[out=180, in=-90] (5.0, -0.7) to[out=90, in=180] (5.8, -0.3)
        -- cycle
    }

    % -----------------------------------------------------
    % RENDER THE SURFACE
    % -----------------------------------------------------
    \shade[left color=blue!45, right color=red!35, even odd rule]
        \outerpath \voidOneA \voidOneB \voidOneC \voidTwoA \voidTwoB ;

    \draw[thick, purple!60!black, even odd rule]
        \outerpath \voidOneA \voidOneB \voidOneC \voidTwoA \voidTwoB ;

    % -----------------------------------------------------
    % 3D PERSPECTIVE ELLIPSES AT THE 5 LEFT OPENINGS
    % -----------------------------------------------------
    \filldraw[fill=blue!20, draw=purple!60!black, thick] (-6, 2.8) ellipse (0.1 and 0.3);
    \filldraw[fill=blue!20, draw=purple!60!black, thick] (-6, 1.4) ellipse (0.1 and 0.3);
    \filldraw[fill=blue!20, draw=purple!60!black, thick] (-6, 0)   ellipse (0.1 and 0.3);
    \filldraw[fill=blue!20, draw=purple!60!black, thick] (-6, -1.4) ellipse (0.1 and 0.3);
    \filldraw[fill=blue!20, draw=purple!60!black, thick] (-6, -2.8) ellipse (0.1 and 0.3);

    % -----------------------------------------------------
    % FOREGROUND AXES & LABELS
    % -----------------------------------------------------
    \draw[->, thick, black!70] (-7.2, -6) -- (12.5, -6) node[right] {$x_5$};
    \draw[->, thick, black!70] (-7.2, -6) -- (-7.2, 6) node[above] {$z$};

    % Brane degree annotations
    \node[above, blue!80!black, font=\bfseries] at (-4.0, 3.7) {$d_0 = 5$};
    \node[above, blue!80!black, font=\bfseries] at (1.6, 2.8)  {$d_1 = 4$};
    \node[above, blue!80!black, font=\bfseries] at (6.8, 2.1)  {$d_2 = 3$};

    \node[align=center, blue!80!black, font=\footnotesize] at (-4.0, 0) {Semi-infinite \\ funnels};
    \node[align=center, blue!80!black, font=\footnotesize] at (1.6, 0) {Suspended \\ segments};
    \node[align=center, blue!80!black, font=\footnotesize] at (6.8, 0) {Suspended \\ segments};

    \end{tikzpicture}
    }
    \caption{The geometric realization of a general recombined D6-brane representing the T-dual of a non-trivial distribution of 5 D5's ending on three D7's accroding to the $[3,1^2]$ partition.}
    \label{fig:t_brane_partition_recombined_543}
\end{figure}
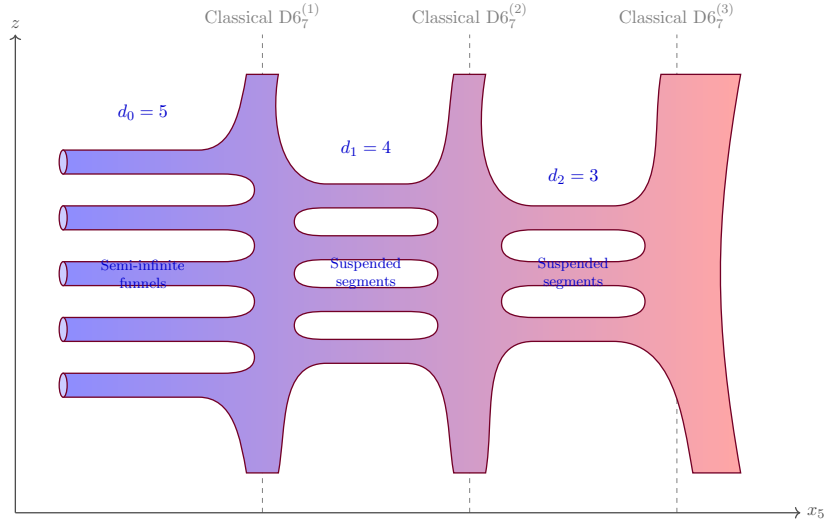

\paragraph{Bringing in 7-branes}
Now we allow each of the D5-branes to end on an individual D7. According to the technology we just developped, the tachyon should take the form
\begin{equation}
    \mathcal{T}_{[1^5]} = 
    \left( \begin{array}{c|c} 
        z\,\mathbf{1}_5 & \mathbf{1}_5 \\ \hline \rule{0pt}{4ex}
        \text{diag}(C_1^{-1}, \dots, C_5^{-1}) & w_-\,\mathbf{1}_5
    \end{array} \right)
\end{equation}
This corresponds to the situation in the following figure \ref{fig:five_staggered_junctions}.
\begin{figure}[h!]
    \centering
    \begin{tikzpicture}[scale=0.7, every node/.style={scale=0.85}]
        
        % Axes for orientation
        \draw[->, thick, gray] (-2.5, -3) -- (-1.5, -3) node[right] {$x_6$};
        \draw[->, thick, gray] (-2.5, -3) -- (-2.5, -2) node[above] {$x_5$};

        % 5 parallel D5-branes and their D7s
        % Ordered: \y (height) / \x (horizontal position of D7)
        % Top is y=2, x=0.8 (leftmost)
        % Bottom is y=-2, x=2.9 (rightmost)
        \foreach \y/\x in {2/0.8, 1/1.3, 0/1.7, -1/2.4, -2/2.9} {
            \draw[thick, blue!80!black] (-1, \y) -- (\x, \y);
            
            % D7-branes
            \draw[thick, fill=white] (\x, \y) circle (0.3);
            % Cross inside D7
            \draw[thick] (\x-0.2, \y-0.2) -- (\x+0.2, \y+0.2);
            \draw[thick] (\x-0.2, \y+0.2) -- (\x+0.2, \y-0.2);
        }
        
        % Labels
        \node[above, blue!80!black] at (0, 2.3) {D5};
        \node[above, font=\footnotesize] at (3, 0.5) {D7's };
        
    \end{tikzpicture}
    \caption{Configuration of 5 parallel D5-branes, each ending on a separate D7-brane. The D7-branes are actually aligned along a horizontal axis, the vertical separation is for illustration purposes only.}
    \label{fig:five_staggered_junctions}
\end{figure}
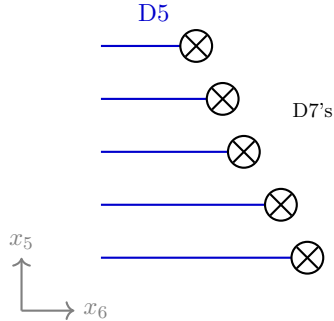
The M-theory uplift of that is
\[
u v = \prod_{i=1}^5 (z-C_i^{-1} w_+)\,.
\]
Next, we activate some T-brane data, for instance
\begin{equation}
    \mathcal{T} = \begin{pmatrix} 
        z\mathbf{1}_5 - \varphi & \mathbf{1}_5 \\ 
        \mathcal{D} + \gamma_1 E_{31}+\gamma_2 E_{54} & w_-\mathbf{1}_5 
    \end{pmatrix}
\end{equation}
where the sub-blocks are defined as:
\begin{itemize}
    \item $\varphi$ is the nilpotent matrix in Jordan form for the $[3, 2]$ partition:
    \[ \varphi = \begin{pmatrix} 0 & 1 & 0 & 0 & 0 \\ 0 & 0 & 1 & 0 & 0 \\ 0 & 0 & 0 & 0 & 0 \\ 0 & 0 & 0 & 0 & 1 \\ 0 & 0 & 0 & 0 & 0 \end{pmatrix} \]
    \item $\mathcal{D} = \text{diag}(C_1^{-1}, \dots, C_5^{-1})$ is the flavor-diagonal coupling matrix.
    \item $E_{i j}$ is the elementary matrix with a unit entry at $(i,j)$ and zeros elsewhere.
\end{itemize}
Switching off $C_1, C_2, C_3$, the M-theory uplift is
\begin{align}
u v &= (z^3-\gamma_1 w_+)\,\left((z-C_3^{-1} w_+)\,(z-C_4^{-1} w_+)-\gamma_2\,w_+ \right)\\
&=-(\gamma_1 C_3^{-1} C_4^{-1}) w_+^3 + \left( z^3 C_3^{-1} C_4^{-1} + \gamma_1 z(C_3^{-1} + C_4^{-1}) \right) w_+^2 - \left( z^4(C_3^{-1} + C_4^{-1}) + \gamma_1 z^2 + \gamma_2 \right) w_+ + z^5
\end{align}
depicted in figure \ref{fig:t_brane_partition_recombined_543}. We could of course switch on further couplings to free up more of the suspended branes here, but this is sufficiently complex to illustrate the mechanisms we are interesting in.
% Requires in your preamble: \usepackage{tikz}, \usetikzlibrary{shadings}, and \usepackage{graphicx} (for \scalebox)

\section{Taub-NUT Geometry and the s-rule}

In the previous sections, we laid the foundation for dualizing simple D5/D7 setups to M-theory geometry. 
Until now, we have restricted our attention to the T-dual of \df-branes and \ds-branes in a flat background. Geometrically, the transverse space in the Type IIA frame was simply the cylinder $\mathbb{C}^* \times \mathbb{C}$, parameterized by the coordinates $(w_\pm, z)$ subject to the trivial fibration constraint $w_+ w_- = 1$. 

To capture the full physics of five-brane webs, specifically the constraints of the s-rule, we must introduce NS5-branes into the Type IIB setup. Let us place a single NS5-brane at the origin of the transverse space, localized at $x_5 = x_7 = x_8 = 0$. Crucially, the NS5-brane does not wrap the $x_9$ circle, but is strictly localized at a point on it. 

When we T-dualize along the $x_9$ direction to reach the Type IIA frame, this localized NS5-brane famously transmutes into pure geometry: a Kaluza-Klein monopole, or a Taub-NUT space. The flat $\mathbb{R}^3 \times S^1$ transverse space deforms such that the radius of the dual T-duality circle $\tilde{x}_9$ shrinks to zero precisely at the location of the original NS5-brane. 

In terms of our complexified algebraic coordinates, the cylinder deforms into a singular conic bundle over the $z$-plane. The defining equation of the Type IIA background becomes:
\begin{equation}\label{eq:taub_nut_bg}
    w_+ w_- = z\,.
\end{equation}
The Taub-NUT center sits exactly at $w_+ = w_- = z = 0$.

In the original Hanany-Witten IIB constructions of 3d theories \cite{Hanany:1996ie}, the so-called naive \emph{s-rule} dictates that no more than one D3-brane may be suspended between an NS5-brane and a D5-brane. Any violation of this breaks supersymmetry. There are several arguments for this, but one very graphical one is that if we suspend $k$ D3-branes between an NS5 and a D5 and shove the D5 past the NS5, 1 D3-brane will be destroyed, and the remaining $k-1$ branes will flip orientation, effectively behaving as anti-D3-branes, not preserving the ambient supersymmetry of the setup. 

The picture is more nuanced for 5-brane webs, as has been discussed in many papers \cite{Arias-Tamargo:2024fjt, Benini:2009gi, Bergman:2012kr, vanBeest:2020kou}. D5 ending on an NS5 isn't quite the correct picture. The D5 actually laminates the NS5, in such a way that half of it is still NS5, and the other half is a $(1,1)$ bound state. This makes the T-duality more subtle. This is briefly but crucially addressed in ABJM \cite{Aharony:2008ug}. Nevertheless, the concept still persists: In a web with $k$ D5-branes, one $(k, 1)$, one $(0,1)$, whereby the $k$ D5's end on a single D7 will break supersymmetry via the same mechanism: Excess anti-branes will appear that will clash with the ambient supersymmetry.

One of the big open questions in the topic of 5-brane web/CY duality is: What is the geometric M-theory counterpart to an s-rule violation?

We will build upon our previous results, and show what this looks like.

\subsection{Building suspended branes} \label{sec:suspendedbranes}
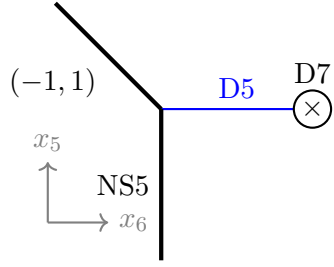
\begin{figure}[h!]
    \centering
    \begin{tikzpicture}[scale=1.0]
        
        % D5-brane (out to the right, shortened to 2)
        \draw[thick, blue] (0,0) -- (2,0) node[midway, above] {D5};
        
        % NS5-brane (out to the bottom, semi-infinite, shortened to -2)
        \draw[ultra thick] (0,0) -- (0,-2) node[midway, left] {NS5};
        
        % (-1,1) brane (out to the top-left, semi-infinite, shortened to -1.4, 1.4)
        \draw[ultra thick] (0,0) -- (-1.4, 1.4) node[midway, below left] {$(-1,1)$};
        
        % ==========================================
        % Draw D7-brane (only on the D5 leg)
        % ==========================================
        
        % D7 Brane (at 2,0, using the standard scalable node format)
        \node[circle, draw, thick, fill=white, inner sep=1pt, minimum size=12pt] (d7) at (2,0) {$\times$};
        \node[above=6pt] at (d7) {D7};
        
        % Axes for orientation (tucked in closer to fit the new bounding box)
        \draw[->, thick, gray] (-1.5, -1.5) -- (-0.7, -1.5) node[right] {$x_6$};
        \draw[->, thick, gray] (-1.5, -1.5) -- (-1.5, -0.7) node[above] {$x_5$};
        
    \end{tikzpicture}
    \caption{The topological vertex in a 5-brane web, where the outgoing D5-brane terminates on a D7-brane at a finite distance. The NS5 and $(-1,1)$ branes extend semi-infinitely.}
    \label{fig:topological_vertex_one_d7}
\end{figure}

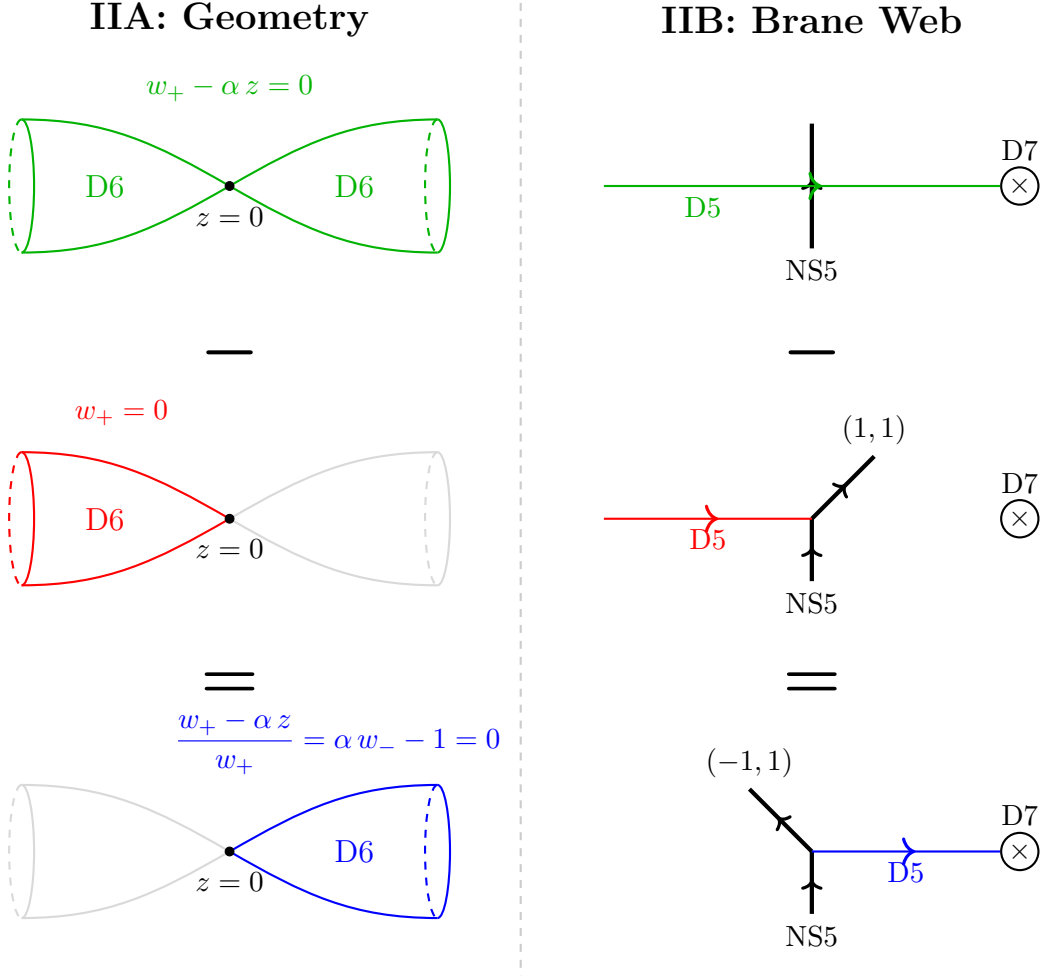
\begin{figure}[h!]
    \centering
    \begin{tikzpicture}[scale=1.1]

        % ==========================================
        % GLOBAL STYLES & MACROS
        % ==========================================
        \tikzset{
            % Standard arrow for 'thick' D5 branes
            ->-/.style={decoration={markings, mark=at position 0.55 with {\arrow[scale=1.5]{>}}},postaction={decorate}},
            % Scaled-down arrow for 'ultra thick' NS5 branes so they match the D5 arrows
            -U-/.style={decoration={markings, mark=at position 0.55 with {\arrow[scale=0.75]{>}}},postaction={decorate}}
        }

        % Reusable macros for drawing the cigars quickly
        \newcommand{\leftcigar}[1]{
            \draw[thick, #1] (0,0) to[out=150, in=0] (-2.5, 0.8);
            \draw[thick, #1] (0,0) to[out=210, in=0] (-2.5, -0.8);
            \draw[thick, #1] (-2.5,-0.8) arc (-90:90:0.15 and 0.8);
            \draw[thick, dashed, #1] (-2.5,0.8) arc (90:270:0.15 and 0.8);
        }
        \newcommand{\rightcigar}[1]{
            \draw[thick, #1] (0,0) to[out=30, in=180] (2.5, 0.8);
            \draw[thick, #1] (0,0) to[out=-30, in=180] (2.5, -0.8);
            \draw[thick, #1] (2.5,-0.8) arc (-90:90:0.15 and 0.8);
            \draw[thick, dashed, #1] (2.5,0.8) arc (90:270:0.15 and 0.8);
        }

        % ==========================================
        % HEADERS & DIVIDER
        % ==========================================
        \node[font=\Large\bfseries] at (-3.5, 2) {IIA: Geometry};
        \node[font=\Large\bfseries] at (3.5, 2) {IIB: Brane Web};
        % Extended vertical line to separate the sides
        \draw[dashed, gray!40, thick] (0, 2.2) -- (0, -9.5);

        % ==========================================
        % ROW 1: THE FULL CONFIGURATION (GREEN)
        % ==========================================
        
        % --- LHS: Full Geometry ---
        \begin{scope}[shift={(-3.5, 0)}]
            \leftcigar{green!70!black}
            \rightcigar{green!70!black}
            \filldraw (0,0) circle(1.5pt) node[below=4pt] {$z=0$};
            
            % Centered the global equation above the whole geometry
            \node[above, text=green!70!black] at (0, .9) {$w_+ - \alpha\,z=0$};
            
            % D6 Labels inside the cigars
            \node[text=green!70!black, font=\large] at (-1.5, 0) {D6};
            \node[text=green!70!black, font=\large] at (1.5, 0) {D6};
        \end{scope}

        % --- RHS: Full D5 to D7 ---
        \begin{scope}[shift={(3.5, 0)}]
            % Active NS5 - Straight vertical (uses the scaled -U- arrow)
            \draw[ultra thick, -U-] (0,-0.75) node[below] {NS5} -- (0,0.75); 
            
            % Active D7
            \node[circle, draw, thick, fill=white, inner sep=1pt, minimum size=12pt] (d7_1) at (2.5,0) {$\times$};
            \node[above=6pt] at (d7_1) {D7};
            
            % Green D5 from -infty to D7 (uses standard ->- arrow)
            \draw[thick, green!70!black, ->-] (-2.5, 0) -- (d7_1) node[pos=0.25, below, text=green!70!black] {D5};
        \end{scope}

        % ==========================================
        % THE MINUS OPERATOR
        % ==========================================
        \node at (-3.5, -2.0) {\Huge $\boldsymbol{-}$};
        \node at (3.5, -2.0) {\Huge $\boldsymbol{-}$};

        % ==========================================
        % ROW 2: THE SUBTRACTED TAIL (RED)
        % ==========================================
        
        % --- LHS: Left Cigar Only ---
        \begin{scope}[shift={(-3.5, -4.0)}]
            \leftcigar{red}
            \rightcigar{gray!30} % Ghosted right cigar
            \filldraw (0,0) circle(1.5pt) node[below=4pt] {$z=0$};
            
            % Label for the left branch
            \node[above, text=red] at (-1.3, 1.0) {$w_+ = 0$};
            
            % D6 Label inside the red cigar
            \node[text=red, font=\large] at (-1.5, 0) {D6};
        \end{scope}

        % --- RHS: D5 to NS5 ---
        \begin{scope}[shift={(3.5, -4.0)}]
            % Active NS5 - Bending 45 deg right (northeast)
            \draw[ultra thick, -U-] (0,-0.75) node[below] {NS5} -- (0,0);
            \draw[ultra thick, -U-] (0,0) -- (0.75, 0.75) node[above] {$(1,1)$};
            
            % SOLID D7
            \node[circle, draw, thick, fill=white, inner sep=1pt, minimum size=12pt] (d7_2) at (2.5,0) {$\times$};
            \node[above=6pt] at (d7_2) {D7};
            
            % Red D5 from -infty to NS5 junction
            \draw[thick, red, ->-] (-2.5, 0) -- (0,0) node[midway, below, text=red] {D5};
        \end{scope}

        % ==========================================
        % THE EQUALS OPERATOR
        % ==========================================
        \node at (-3.5, -6.0) {\Huge $\boldsymbol{=}$};
        \node at (3.5, -6.0) {\Huge $\boldsymbol{=}$};

        % ==========================================
        % ROW 3: THE RESULTING SUSPENDED BRANE (BLUE)
        % ==========================================
        
        % --- LHS: Right Cigar Only ---
        \begin{scope}[shift={(-3.5, -8.0)}]
            \leftcigar{gray!30} % Ghosted left cigar
            \rightcigar{blue}
            \filldraw (0,0) circle(1.5pt) node[below=4pt] {$z=0$};
            
            % Equation
            \node[above, text=blue] at (1.3, .8) {$\displaystyle \frac{w_+-\alpha\,z}{w_+} = \alpha\,w_- - 1 = 0$};
            
            % D6 Label inside the blue cigar
            \node[text=blue, font=\large] at (1.5, 0) {D6};
        \end{scope}

        % --- RHS: Suspended D5 ---
        \begin{scope}[shift={(3.5, -8.0)}]
            % Active NS5 - Bending 45 deg left (northwest)
            \draw[ultra thick, -U-] (0,-0.75) node[below] {NS5} -- (0,0);
            \draw[ultra thick, -U-] (0,0) -- (-0.75, 0.75) node[above] {$(-1,1)$};
            
            % Active D7
            \node[circle, draw, thick, fill=white, inner sep=1pt, minimum size=12pt] (d7_3) at (2.5,0) {$\times$};
            \node[above=6pt] at (d7_3) {D7};
            
            % Blue Suspended D5
            \draw[thick, blue, ->-] (0, 0) -- (d7_3) node[midway, below, text=blue] {D5};
        \end{scope}

    \end{tikzpicture}
    \caption{Visualization of the subtraction process. Left (IIA): The fully reducible D6-brane on the Taub-NUT geometry minus the $w_+=0$ branch leaves exactly the finite $\alpha w_--1=0$ branch. Right (IIB): The T-dual of a semi-infinite D5-brane stretching to the D7, minus the semi-infinite segment stretching to the NS5, isolating exactly the suspended D5-brane segment.}
    \label{fig:full_subtraction}
\end{figure}

\paragraph{One suspended brane}
First, we explain how to build the T-dual of a suspended D5-brane as in figure \ref{fig:topological_vertex_one_d7}.
We will do this step by step:
\begin{itemize}
\item First, consider a D5-brane coming from the infinite left, and ending on a D7 without passing through the NS5. Its T-dual can be readily defined via the following Ansatz
\[ w_+-\alpha\,(z-z_0)=0\,.\]
To see this, we notice that, for $z_0 \neq 0$, this D6-brane never hits the TN center at $\{w_\pm = 0 \cap z=0\}$. Hence, we can work in a local patch where the ambient geometry looks like $\mathbb{C}^* \times \mathbb{C}$, and all the results from the previous sections apply.

\item Now we tune $z_0 \rightarrow 0$, and we immediately notice that our brane becomes reducible. In terms of ideals:
\[ (w_+-\alpha\,z; \, w_+\,w_--z)\quad \cong \quad (w_+-\alpha\,z; \, w_+\,(\alpha w_--1)\,. \]
We can now interpret what is happening as follows: On the IIB side, we have slid the semi-infinite D5 along the D7's $z$-plane, all the way to its origin. The D5 now intersects the NS5, which is located at $(x_6=0, z=0)$, and can now break into two halves: One semi-infinite D5 coming from the far left ending on the NS5, and one suspended D5-segment between the NS5 and the D7\footnote{We remind the reader that this is an oversimplification. As explained before, the D5 never really 'ends' on an NS5, but actually merges along it, creating a $(1,1)$ 5-brane in one half of the $x_6$ axis, which leaving a pure NS5 in the other half. Nevertheless, we will not be severly punished for this sloppiness.}.
On the T-dual side, this splitting translates to the reducibility of our Riemann surface. The $(w_+-\alpha\,z; \, w_+) \cong (z; w_+)$ branch is the T-dual of the semi-infinite D5. It wraps the leftmost cigar in \ref{fig:full_subtraction}. Hence, the remaining branch $(w_+-\alpha\,z; \, \alpha w_--1)$ is the T-dual of the suspended brane.
\end{itemize}
The algebraic geometry reveals something interesting, the equation 
\begin{equation} \label{eq:1suspendedbrane}
    \alpha\,w_--1 = 0
\end{equation} 
can simply be recast as
\begin{equation} \label{eq:hwtransitioned1suspendedbrane}
    (w_+-\alpha\,z)/(w_+)=0\,.
    \end{equation}
Equation \eqref{eq:hwtransitioned1suspendedbrane} shows us what we were trying to build, the T-dual of a suspended D5, as a quotient. We can think of this as a divisor of a rational function $f$, following the rule
\begin{equation}
    {\rm Div}(f) = [{\rm{zeroes}(f)}] - [{\rm{poles}(f)}]\,.
\end{equation}
We are taking a D6 that corresponds to a D5 going straight from the infinite left all the way to the D7, and subtracting from it a semi-infinite D5 that goes from the infinite left to the NS5. This is schematically shown in figure \ref{fig:full_subtraction}. The numerator is the T-dual of a semi-infinite D5 stretching all the way from the left, past the NS5, ending on a D7 at finite distance to the right of the NS5. The denominator is the T-dual of a semi-infinite D5 stretching from the left up to the NS5 and stopping there. The rational function \eqref{eq:hwtransitioned1suspendedbrane} gives the homological difference between these two, representing exactly the brane we need.
The M-theory uplift in this framing is
\[
u\,v =     (w_+-\alpha\,z)/(w_+)\,, \qquad \cap \qquad w_+\,w_-=z
\]
The second presentation \eqref{eq:1suspendedbrane} seems to show the T-dual of a D7, with no D5's at all. This is actually the Hanany-Witten transition at play, in this $w_-$-frame, the D7 lies to the left of the NS5, and the D5 has been destroyed. 
The M-theory uplift of this is
\[
u\,v =     \alpha\,w_--1\,, \qquad \cap \qquad w_+\,w_-=z\,.
\]
After eliminating the $z$-variable, we see a perfectly smooth geometry. The M-theory circle collapses at $w_-=1/\alpha$, signaling the T-dual of the D7-brane.

\begin{figure}[h!]
    \centering
    \begin{tikzpicture}[scale=1.2]
        
        % ==========================================
        % LEFT PANEL: Unwarped Plane
        % ==========================================
        \begin{scope}[shift={(-3.5,0)}]
            
            % TN Center (placed firmly on the right)
            \filldraw[black] (.8, 0.5) circle (2pt);
            \node[right=2pt] at (.8, 0.5) {TN center};
            
            % D6_7 Plane (Flat, strictly to the left)
            \filldraw[fill=blue!10, draw=blue, thick, join=round] 
                (-1.5, -2.5) -- (-0.5, -1.5) -- (-0.5, 3.5) -- (-1.5, 2.5) -- cycle;
            
            % Vertical mesh lines (baseline)
            \foreach \x in {-1.35, -1.2, -1.05, -0.9, -0.75, -0.6} {
                \draw[blue!40, thin] (\x, {\x+4.0}) -- (\x, {\x-1.0});
            }
            
            % Horizontal transverse mesh lines (matching the right panel's perspective)
            \begin{scope}
                \clip (-1.5, -2.5) -- (-0.5, -1.5) -- (-0.5, 3.5) -- (-1.5, 2.5) -- cycle;
                \foreach \y in {-2.0, -1.5, -1.0, -0.5, 0, 0.5, 1.0, 1.5, 2.0, 2.5, 3.0} {
                    \draw[blue!30, very thin] (-2, \y) -- (0, \y);
                }
            \end{scope}
            
            % Label for the main plane
            \node[blue!80!black] at (-1, 3.2) {$\text{D6}_7$};
            
            % Subcaption
            \node at (0, -3.5) {(a) $\alpha\,w_--1=0$};
            
        \end{scope}

        % ==========================================
        % RIGHT PANEL: Elongated Crevasse (D6_5 Tube)
        % ==========================================
        \begin{scope}[shift={(3.5,0)}]
            
            % 1. Solid background fill to define the continuous sheet
            \fill[blue!10]
                (2.5, 3.5) -- (1.5, 2.5)
                % Left profile: Elongated heavily to x = -2.0
                -- (1.5, 1.7)
                .. controls (1.5, 1.1) and (-2.0, 1.1) .. (-2.0, 0.5)
                .. controls (-2.0, -0.1) and (1.5, -0.1) .. (1.5, -0.7)
                -- (1.5, -2.5) -- (2.5, -1.5) -- cycle;

            % 2. Draw Top and Bottom edges explicitly
            \draw[blue, thick] (2.5, 3.5) -- (1.5, 2.5);
            \draw[blue, thick] (2.5, -1.5) -- (1.5, -2.5);

            % 3. DENSE TRANSVERSE MESH (Horizontal Lines)
            % Using a clip makes the lines perfectly conform to the 3D topology
            \begin{scope}
                \clip 
                    (2.5, 3.5) -- (1.5, 2.5) -- (1.5, 1.7)
                    .. controls (1.5, 1.1) and (-2.0, 1.1) .. (-2.0, 0.5)
                    .. controls (-2.0, -0.1) and (1.5, -0.1) .. (1.5, -0.7)
                    -- (1.5, -2.5) -- (2.5, -1.5) -- cycle;
                
                \foreach \y in {-2.0, -1.7, -1.4, -1.1, -0.8, -0.5, -0.2, 0.1, 0.4, 0.7, 1.0, 1.3, 1.6, 1.9, 2.2, 2.5, 2.8, 3.1} {
                    \draw[blue!30, very thin] (-2.5, \y) -- (2.5, \y);
                }
                % Emphasize the exact equator line of the tube
                \draw[blue!40, thin] (-2.5, 0.5) -- (2.5, 0.5);
            \end{scope}

            % 4. VERTICAL MESH LINES (Waterfall effect stretched to x = -2.0)
            % Removed the padding spaces and the \ifx conditionals to fix the compilation error
            \foreach \x/\tipx/\col/\thickness in {
                2.5/2.5/blue/thick,
                2.4/2.455/blue!50/thin,
                2.25/2.218/blue!50/thin,
                2.1/1.78/blue!50/thin,
                1.95/1.138/blue!50/thin,
                1.8/0.3/blue!50/thin,
                1.65/-0.75/blue!50/thin,
                1.5/-2.0/blue/thick%
            } {
                \draw[\col, \thickness] 
                    (\x, {\x+1.0}) -- (\x, 1.7)
                    .. controls (\x, 1.1) and (\tipx, 1.1) .. (\tipx, 0.5)
                    .. controls (\tipx, -0.1) and (\x, -0.1) .. (\x, -0.7)
                    -- (\x, {\x-4.0});
            }

            % 5. LABELS: D6_7 and D6_5
            \node[blue!80!black] at (2, 3.2) {$\text{D6}_7$};
            % Placed halfway along the stretched horizontal tube
            \node[blue!90!black, fill=blue!10, inner sep=1.5pt, rounded corners=2pt] 
                at (-0.5, 1.25) {$\text{D6}_5$};

            % 6. TN Center (Positioned to the right of the extreme apex)
            % The apex is at x = -2.0. The TN center is shifted right to x = -1.75.
            \filldraw[black] (-1.75, 0.5) circle (2pt);
            \node[right=2pt, fill=white, fill opacity=0.7, text opacity=1, inner sep=1pt, rounded corners=2pt] 
                at (-1.75, 0.5) {TN center};
            
            % Subcaption
            \node at (0.5, -3.5) {(b) $\frac{w_+-\alpha\,z}{w_+}=0$};
            
        \end{scope}

    \end{tikzpicture}
    \caption{Left: A flat $\text{D6}_7$-brane localized to the left of a TN-center. Right: The same brane in $w_+$-frame, now seen as trying to cross the TN-center. As the brane moves right, the surface is continuously deformed, extruding a long horizontal tube that represents the $\text{D6}_5$-brane. This tube perfectly wraps around the Taub-NUT center without intersecting it, and corresponds to the T-dual of the suspended D5-brane.}
    \label{fig:elongated_tube}
\end{figure}
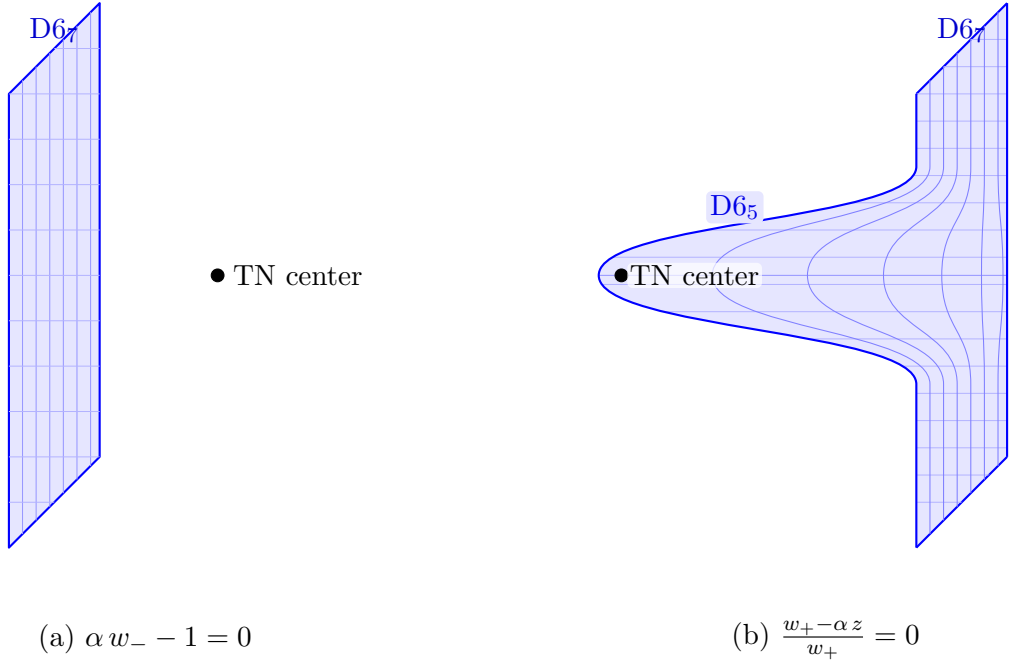

What is curious in this construction, is that the HW transitions are implemented, not by changing the moduli of the geometry or the branes, but by changing coordinate frames. This is suggestive of a picture whereby brane creation is really nothing but the formation of elongated bend branes as depicted in figure \ref{fig:elongated_tube}

We can deduce the form \eqref{eq:hwtransitioned1suspendedbrane} from the tachyon condensation picture as follows: Start with a D6$_5$ brane at $w_+=0$. This is the T-dual of a semi-infinite D5 going from $x_5=-\infty$ to the NS5 at $x_5=0$. We can write down its tachyon describe, albeit redundantly, as follows:
\[
\begin{pmatrix}
    w_+ & 0\\0& 1\,,
\end{pmatrix}
\]
where we have added a harmless trivial direct summand. Now activate some T-brane data as follows:
\[
\begin{pmatrix}
    w_+ & \alpha\\z& 1\,.
\end{pmatrix}
\]
This now describes a brane at $w_+-\alpha\,z$, which is the T-dual to a semi-infinite brane from the far left to the D7, as explained before. Now, we can bring this matrix in Schur complement form as follows:
\[
\begin{pmatrix}
    w_+ & 0\\0& \frac{w_+-\alpha\,z}{w_+}\,.
\end{pmatrix}
\]
Now, we can extract the $(1,1)$ block away via the \emph{cone construction} in homological algebra. This corresponds to adding an anti-brane at $w_+$ and attaching it to this block, triggering annihilation. The leftover will be the desired brane. See the appendix \ref{app:schur} for more detail on the relation between Schur complements and the cone construction.

\paragraph{Multiple suspended branes:} We can leverage our geometric understanding of HW transitions to build the case for the M-theory of multiple D5's suspended between the NS5 and muliple D7's. We start with the M-theory dual of $k$ D7-branes to the left of the NS5:
\begin{equation} \label{eq:srulerespectnopoles}
    u\,v = \prod_{i=1}^k (\alpha_i\,w_--1)\,, \qquad \cap \qquad w_+\,w_-=z\,,
\end{equation}
and recast it in $w_+$-frame as follows
\begin{equation} \label{eq:srulerespectwithpoles}
    u\,v = \frac{1}{w_+^k}\,\prod_{i=1}^k (z-\alpha_i\,w_+)\,, \qquad \cap \qquad w_+\,w_-=z\,,
    \end{equation}
This displays the HW transition very nicely again. The presentation in \eqref{eq:srulerespectnopoles} shows us the T-dual of k unattached D7-branes to the left of the NS5/(-k,1) junction, whereas the framing in \eqref{eq:srulerespectwithpoles} shows us the T-dual of k D7's to the right of the NS5/(-k, 1), each with a suspended D5 ending on it.

\subsection{S-rule violation} \label{sec:sviol}
\begin{figure}
    \centering
    % Resize the first diagram to fit the text width
    \resizebox{\textwidth}{!}{%
    \begin{tikzpicture}[
        arrowRight/.style={decoration={markings, mark=at position 0.5 with {\arrow{>}}},postaction={decorate}},
        arrowLeft/.style={decoration={markings, mark=at position 0.5 with {\arrow{<}}},postaction={decorate}},
        arrowU/.style={decoration={markings, mark=at position 0.6 with {\arrow{>}}},postaction={decorate}}
    ]
        % ==================================================
        % FIGURE 1: w_+ Frame Subtraction
        % ==================================================
        \node[font=\bfseries] at (6.5, 2.5) {Subtraction in $w_+$-frame ($uv = \frac{w_+ - \alpha z^k}{w_+^k}$)};
        
        % Row 1: k D5 to D7
        \draw[ultra thick, arrowU] (0,-1) node[below] {NS5} -- (0,0) -- (0,1) ;
        \draw[thick, blue, arrowRight] (-2.5, 0) -- (2, 0) node[pos=0.3, above] {$k$ D5};
        \node[circle, draw, thick, fill=white] at (2,0) {$\times$};
        
        \node at (3.5, 0) {\Large $-$};
        
        % Row 2: k Semi-infinite D5s to NS5
        \begin{scope}[shift={(7,0)}]
            \draw[ultra thick, arrowU] (0,-1) node[below] {NS5} -- (0,0) -- (0.7, 0.7) node[above] {$(k,1)$};
            \draw[thick, red, arrowRight] (-2.5, 0) -- (0, 0) node[midway, below] {$k$ D5};
        \end{scope}

        \node at (10, 0) {\Large $=$};

        % Row 3: Result (k Suspended D5s)
        \begin{scope}[shift={(13,0)}]
            \draw[ultra thick, arrowU] (0,-1) node[below] {NS5} -- (0,0) -- (-0.7, 0.7) node[above] {$(-k,1)$};
            \draw[thick, green!60!black, arrowRight] (0, 0) -- (2, 0) node[midway, above] {$k$ D5};
            \node[circle, draw, thick, fill=white] at (2,0) {$\times$};
        \end{scope}
    \end{tikzpicture}
    }

    \vspace{1cm}

    % Resize the second diagram to fit the text width
    \resizebox{\textwidth}{!}{%
    \begin{tikzpicture}[
        arrowRight/.style={decoration={markings, mark=at position 0.5 with {\arrow{>}}},postaction={decorate}},
        arrowLeft/.style={decoration={markings, mark=at position 0.5 with {\arrow{<}}},postaction={decorate}},
        arrowU/.style={decoration={markings, mark=at position 0.6 with {\arrow{>}}},postaction={decorate}}
    ]
        \node[font=\bfseries] at (3.5, 2.5) {Subtraction in $w_-$-frame ($uv = \frac{1 - \alpha z^{k-1} w_-}{w_+^{k-1}}$)};
        
        % Row 1: k-1 D5 to D7
        \draw[ultra thick, arrowU] (-1,-1) node[below] {NS5} -- (-1,0) -- (-1,1) node[above] {$(0,1)$};
        \node[circle, draw, thick, fill=white] (d7) at (-2.5,0) {$\times$};
        \draw[thick, blue, arrowRight] (-4,0) -- (d7) node[pos=0.3, above] {$k-1$ D5};
        
        \node at (1.5, 0) {\Large $-$};
        
        % Row 2: k-1 Semi-infinite D5s to NS5
        \begin{scope}[shift={(4.5,0)}]
            \draw[ultra thick, arrowU] (0,-1) node[below] {NS5} -- (0,0) -- (0.7,0.7) node[above] {$(1,1)$};
            \draw[thick, red, arrowRight] (-2.5, 0) -- (0,0) node[midway, below] {$k-1$ D5};
        \end{scope}

        \node at (6.0, 0) {\Large $=$};

        % Row 3: Result (k-1 suspended anti-D5s)
        \begin{scope}[shift={(9,0)}]
            \draw[ultra thick, arrowU] (0,-1) node[below] {NS5} -- (0,0) -- (-0.7,0.7) node[above] {$(-1,1)$};
            \node[circle, draw, thick, fill=white] (d7) at (-2,0) {$\times$};
            \draw[thick, red, arrowLeft] (d7) -- (0,0) node[midway, below] {$k-1$ $\overline{\text{D5}}$};
        \end{scope}
    \end{tikzpicture}
    }
    \caption{Subtraction schemes. (Top) Construction of suspended branes in the $w_+$-frame. (Bottom) HW transition to the $w_-$-frame: $k-1$ semi-infinite D5s to D7 minus $k-1$ semi-infinite D5s to NS5 leaves $k-1$ suspended anti-D5 branes.}
    \label{fig:sviol}
\end{figure}
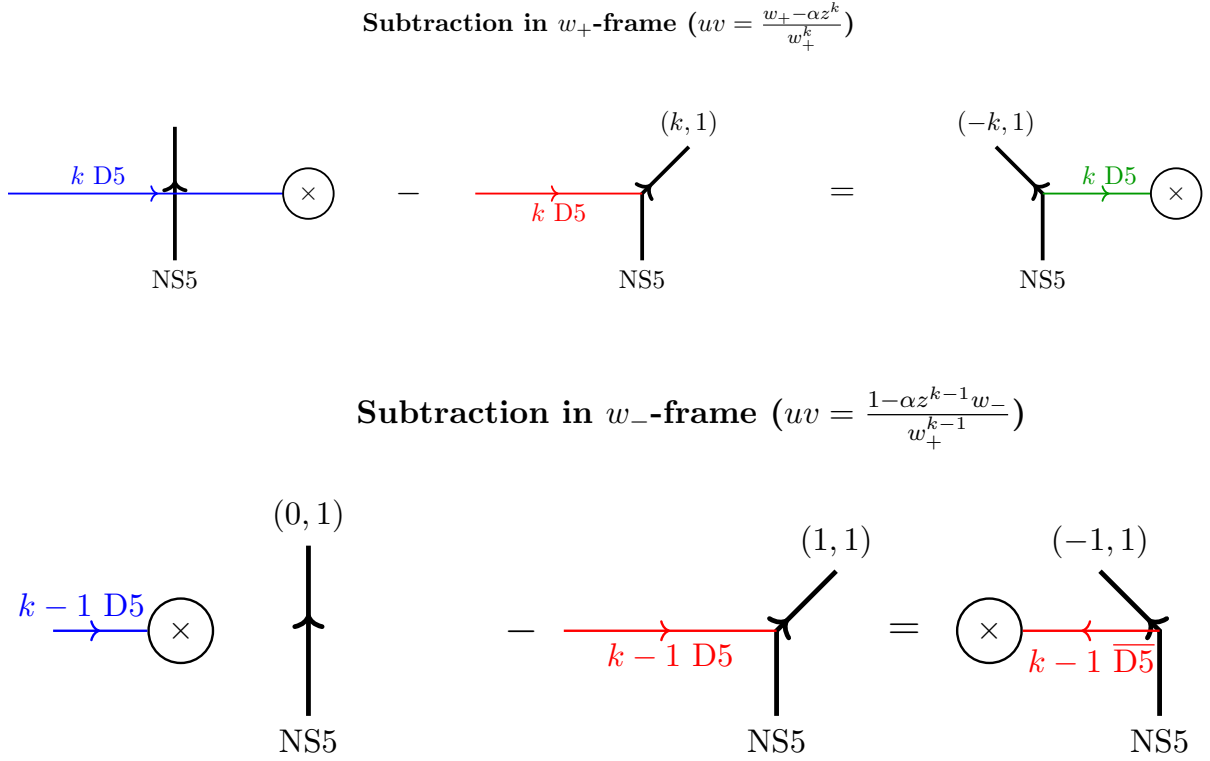
Let us now construct a propotype of the s-rule violation. On the IIB side, the simplest example is having $k>1$ D5-branes suspended between a single D7 and a single NS5. This simply amounts to generalizing our previous results. The M-theory geometry is simply:
\[u v = \frac{w_+-\alpha\,z^k}{w_+^k}\,.\]
The logic is the same as before, the rhs describes a subtraction of divisors, such that we end up with exactly the suspended branes we want, and eliminate the T-dual of the excess semi-infinite part. Now let's see what happens when we pass to $w_-$-frame:
\[u v = \frac{1-\alpha\,z^{k-1}\,w_-\,\,}{w_+^{k-1}}\,.\]
This equation has a beautiful physical interpretation. It corresponds exactly to what one would expect from a HW transition for an s-rule violating configuration. The numerator corresponds to the T-dual of $k-1$ semi-infinte D5-branes from the far left to a D7, lying to the left of the NS5/$(-k,1)$ 5-brane junction.
The denominator subtracts $k-1$ semi-infinite D5-branes going from the far left to the NS5. The net result: $k-1$ \emph{anti}-D5-branes suspended between the D7 and the NS5. Both frames are summarized in figure \ref{fig:sviol}.

How can we diagnose the s-rule violation from a purely geometrical perspective? Here, the smoking gun is that there is no framing of this geometry that clears poles. Let's contrast this to the situation in the previous section, where the setup that appeared to have poles in \eqref{eq:srulerespectwithpoles}, could be recast into a pole-free situation \eqref{eq:srulerespectnopoles}. There, the M-theory circle remains finite over the entire base of the geometry. 
On the contrary, in the present s-rule violating configuration, there is no coordinate system that gets rid of the poles. They are not an artifact. The M-theory circle blows up over points at a finite distance in the base geometry.

\paragraph{Comment on anti-D6-branes:}
We do no expect this to be the full physical picture: The M-theory uplift of an anti-D6-brane is an anti-Taub-NUT geometry, which means that the $S^1$ fibration has a negative Chern class, when restricted to a $U(1)$-bundle over a linking 2-sphere. In any case, and anti-D6-brane has positive tension.
At the level of topological B-branes, we can think of anti-branes in the derived category of coherent sheaves as regular branes, with a negative coefficient in homology, and the na\"ive picture of brane/anti-brane annihilation goes through. We are well aware, however, that this isn't the whole story. A D6/anti-D6 pair do not simply annihilate into pure nothingness. A slew of closed strings will be emitted. M-theory is supposed to capture the full backreaction, and hence should be sensitive to this information. It is therefore intriguing that the prescription:
\[
u v = [{\rm zeroes}(f)]-[{\rm poles}(f)]
\]
seems to give the right picture.

\section{Multiple legs}
In the previous sections, we learned how to deform CY geometries, in accordance to what happens when we alter the behavior of \emph{one} asymptotic leg of a given 5-brane web diagram. We achieved this by going to a duality frame where that leg is made entirely of perturbative D5/D7 branes, and the NS5/(-k,1) junction dualizes to pure TN geometry in IIA. However, the methods developed here can be applied patchwise on any CY threefold.

In this section, we will setup the general procedure, and then specialize to the class of examples often referred to as $T_N$ theories, which correspond to M-theory on $\mathbb{C}^3/(\mathbb{Z}_N \times \mathbb{Z}_N )$. Their toric graphs are equilateral triangles with all three sides of length $N$. 

\subsection{General procedure}
A toric CY threefold can be described via a 2d polytope that corresponds to the `ceiling' of its 3d fan. Concretely, all fan generators are vertices of height one.
By triangulating such a polytope, one encodes the various possible K\"ahler resolutions of the threefold singularity. However, in this paper, we will keep the singularity in the spirit of describing a 5d SCFT, and \emph{not} triangulation the fan. How does one then describe such a variety from the fan? One defines its \emph{toric ideal}. The procedure is reviewed in many textbooks, such as \cite{fulton1993introduction}. It goes roughly as follows:
Given a maximal three-dimensional cone $\sigma_3 = \langle v_1\, \ldots\, v_n \rangle$ spanned by the height-one generators $v_i \in M$ in the lattice $M$, one defines its dual cone in the dual lattice $N$, as follows
\[
\sigma_3^\vee: =\{m \in M| \, \langle m, v \rangle \geq 0, \forall v \in \sigma_3 \}
\]
Now, $\sigma_3^\vee = \langle m_1\,\ldots\, m_k \rangle$ will be spanned by several 'monomials' $m_i$. A theorem guarantees that these monomials satisfy a number of binomial relations Rel$_a$ of the form 
\[{\rm Rel_a}: \sum_{i_a^+} m_{i_a^+} - \sum_{i_a^-} m_{i_a^-}  = 0\,.
\]

This will give rise to a (typically) non-complete intersection. This complete intersection of binomials cuts out the affine, singular CY geometry.

The added advantage of this description, is that it is possible to specialize to patches. In our case, we want to understand what happens at the vicinity of each external leg of the 5-brane web. What does this correspond to in the toric description? It corresponds to focusing on one edge of the polytope, say, the one spanned by $\langle v_i\, v_{i+1} \rangle$, where $v_i$ and $v_{i+1}$ are the extremal vertices of the edge. If the 2-cone has length $l_i$, the there will be $l_i-1$ further vertices in interior of the edge. But those will not interest us. This length $l_i$ edge corresponds to a two-dimensional cone in the fan. One can define the two-cone $\tau_{i, i+1}$, and its dual cone $\tau^\vee_{i, i+1}$. It actually corresponds to an patch of our toric CY threefold. Just as for the 3-cone, one can establish a toric ideal for it. So, we define
\[
U_{\tau_{i, i_1}} = {\rm Spec}\mathbb{C}[\tau^\vee_{i, i+1}]\,.
\]
Since these are all two-cones, they necessarily all take the form of a local toric K3 times a cylindrical transverse direction. More precisely, they all have the local structure:
\begin{equation}
U_{\tau_{i, i_1}} = {\rm Spec}\mathbb{C}[u_i, v_i, z_i, s_i^\pm]/(u_i v_i-z_i^{l_i})\,,
\end{equation}
corresponding to $\mathbb{C}^2/\mathbb{Z}_{l_i} \times \mathbb{C}^*$.

The full CY threefold can be reconstructed by gluing together several such open sets
\[
\bigcup_{\rm edges}\, \mathbb{C}^2/\mathbb{Z}_{l_i} \times \mathbb{C}^*\,.
\]
In each such patch, one can implement the deformations described in the previous sections. Determining global consistency then requires checking that they glue together correctly. However, we have not yet found examples where there are clashes leading to s-rule violations. Rather, we found that the s-rule gets violated already by activating T-brane data on even a single edge (if conditions trigger it), as in the case discussed in section \ref{sec:sviol}.

\subsection{Examples}
\subsubsection{Deforming $\mathbb{C}^3$}
Building upon the result in section \ref{sec:suspendedbranes}, we now apply this to the three patches of $\mathbb{C}^3$. For this case, the original toric ideal is simple:
\[
C[X_1, X_2, X_3, W]/(X_1 X_2 X_3-W)\,.
\]
This description has a redundant coordinate $W$. The toric polytope is an equilateral triangle with all sides of size 1, corresponding to a simple `topological vertex' like 5-brane web. There are three possible 2-cones with which to build patches that are isomorphic to $\mathbb{C}^2 \times \mathbb{C}^*$. The are defined as follows:
\begin{itemize}
    \item Patch 1: $X_1 \neq 0$. Its coordinate ring is $R_1 = \mathbb{C}[u_1, v_1, z_1, s_1^\pm]/(u_1 v_1-z_1)$ with 
    \[
    u_1:=X_1 X_2\,, \quad v_1:= X_3\,,\quad z_1 := W\,,\quad s_1:= X_1\,.
    \]
    \item Patch 2: $X_2 \neq 0$. Its coordinate ring is $R_1 = \mathbb{C}[u_2, v_2, z_2, s_2^\pm]/(u_2 v_2-z_2)$ with 
    \[
    u_2:=X_1 X_2\,, \quad v_2:= X_3\,,\quad z_2 := W\,,\quad s_2:= X_2\,.
    \]
    \item Patch 3: $X_3 \neq 0$. Its coordinate ring is $R_1 = \mathbb{C}[u_3, v_3, z_3, s_3^\pm]/(u_3 v_3-z_3)$ with 
    \[
    u_3:=X_1\,, \quad v_3:= X_2 X_3\,,\quad z_3 := W\,,\quad s_3:= X_3\,.
    \]
\end{itemize}
Now we can deform each one by applying the logic in \eqref{eq:simplejunction}. For instance we can deform patch 1 via
\[
u_1 v_1 = z_1+\alpha_1 s_1
\]
which, in global coordinates, amounts to
\[
X_1 X_2 X_3 = W+\alpha_1 X_1\,.
\]
What does this deformation look like in, say, patch 2? By applying the appropriate transformations, we find it becomes
\[
u_2 v_2 = z_2+\alpha_1 \frac{u_2}{s_2}\,.
\]
Two things to note: Firstly, the pole is not problematic, since $s_2 \in \mathbb{C}^*$. Secondly, notice that this deformation lies in the local Jacobian ideal
\[J_2 = \vec{\nabla} (u_2 v_2-z_2) = (v_2, u_2, -1)\,.
\]
In other words, it does not correspond to a versal deformation, and can be reabsorbed by a shift in the coordinate $v_2 \rightarrow v_2-\alpha_1/s_2$\,. So the D7 appearing on one leg of the vertex seems to have no effect on the other legs.

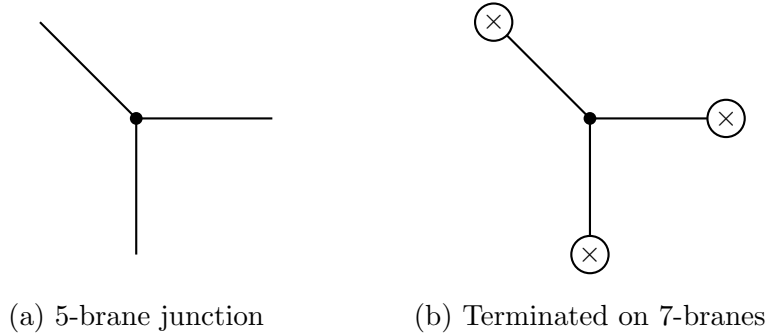
\begin{figure}[htbp]
    \centering
    \begin{tikzpicture}[thick, scale=1.2]

        % ---------- Panel 1: Topological Vertex ----------
        \begin{scope}
            \coordinate (O) at (0,0);
            \draw (O) -- (135:1.5);
            \draw (O) -- (270:1.5);
            \draw (O) -- (0:1.5);
            \fill (O) circle (2pt);
            \node at (0,-2.2) {(a) 5-brane junction};
        \end{scope}

        % ---------- Panel 2: Terminated on 7-branes ----------
        \begin{scope}[xshift=5cm]
            \coordinate (O) at (0,0);
            \coordinate (NW) at (135:1.5);
            \coordinate (D)  at (270:1.5);
            \coordinate (R)  at (0:1.5);

            \draw (O) -- (NW);
            \draw (O) -- (D);
            \draw (O) -- (R);
            \fill (O) circle (2pt);

            \node[circle, draw, fill=white, inner sep=1pt] at (NW) {$\times$};
            \node[circle, draw, fill=white, inner sep=1pt] at (D)  {$\times$};
            \node[circle, draw, fill=white, inner sep=1pt] at (R)  {$\times$};

            \node at (0,-2.2) {(b) Terminated on 7-branes};
        \end{scope}

    \end{tikzpicture}
    \caption{Two configurations of 5-brane junctions: (a) the pure 5-brane junction (b) the same web with legs terminating on 7-branes.} \label{fig:deformingc3}
\end{figure}

Therefore, we can freely superpose such deformations, leading to
\[
X_1 X_2 X_3 = W+\alpha_1 X_1+\alpha_2 X_2+\alpha_3 X_3\,.
\]
We claim that this is the effect of bringing in three 7-branes from infinitely far away to within finite distance as in figure \ref{fig:deformingc3}.

Topologically, nothing has happened here, but clearly at the level of the complex structure, something important may be at play. On the IIB side, having brought in the 7-branes to a finite distance effectively compactifies the 5-brane junction. This means that it's fluctuations are not six-dimensional any more, but might be genuine five-dimensional degrees of freedom. It would be interesting to elucidate this point by different methods.\footnote{I thank Michele Del Zotto for discussions about this intriguing and unexpected question.}

\subsubsection{Deforming $T_N$}
\begin{figure}[htbp][h!]
    \centering
    \begin{tikzpicture}[scale=1.2, baseline=(current bounding box.center)]
        % Styles
        \tikzset{
            brane7/.style={circle, draw, thick, fill=white, inner sep=1.5pt}
        }

        % ==========================================
        % NW Leg: Partition [1^3] (3 separate 7-branes)
        % ==========================================
        \coordinate (NW1) at (-1.7, 1.9);
        \coordinate (NW2) at (-1.8, 1.8);
        \coordinate (NW3) at (-1.9, 1.7);
        
        \draw[thick] (0.1, 0.1) -- (NW1);
        \draw[thick] (0, 0) -- (NW2);
        \draw[thick] (-0.1, -0.1) -- (NW3);
        
        \node[brane7] at (NW1) {$\times$};
        \node[brane7] at (NW2) {$\times$};
        \node[brane7] at (NW3) {$\times$};
        
        \node[above right] at (-1.2, 1.2) {$3$ D5};
        \node[above left] at (NW1) {$[1^3]$};

        % ==========================================
        % Down Leg: Partition [2,1] (2 branes merge, 1 separate)
        % ==========================================
        \coordinate (D1) at (-0.075, -2.5); % Merged 7-brane
        \coordinate (D2) at (0.15, -2.5);   % Single 7-brane
        
        \draw[thick] (-0.15, 0) -- (-0.15, -2.0) -- (D1);
        \draw[thick] (0, 0) -- (0, -2.0) -- (D1);
        \draw[thick] (0.15, 0) -- (0.15, -2.5);
        
        \node[brane7] at (D1) {$\times$};
        \node[brane7] at (D2) {$\times$};
        
        \node[right] at (0.15, -1.2) {$3$ D5};
        \node[below] at (0.03, -2.8) {$[2,1]$};

        % ==========================================
        % Right Leg: Partition [3] (All 3 branes merge)
        % ==========================================
        \coordinate (Rend) at (2.5, 0);
        
        \draw[thick] (0, 0.15) -- (2.0, 0.15) -- (Rend);
        \draw[thick] (0, 0) -- (Rend);
        \draw[thick] (0, -0.15) -- (2.0, -0.15) -- (Rend);
        
        \node[brane7] at (Rend) {$\times$};
        
        \node[above] at (1.2, 0.15) {$3$ D5};
        \node[right] at (2.7, 0) {$[3]$};

        % ==========================================
        % Central Junction
        % ==========================================
        % A filled circle large enough to cover the initial offsets
        \fill (0,0) circle (7pt);
        
    \end{tikzpicture}
    \caption{A triple junction of 5-branes ($k=3$) terminating on 7-branes according to specific partitions. The Northwest leg terminates fully on distinct 7-branes ($[1^3]$), the Down leg partially condenses ($[2,1]$), and the Right leg fully condenses into a single 7-brane ($[3]$).} \label{fig:T3}
\end{figure}
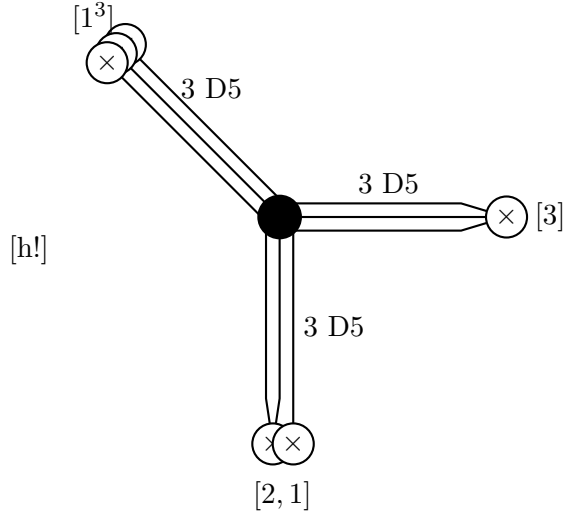

We can easily generalize this to the non-Abelian case. In \cite{Bourget:2023wlb}, the deformations were deduced by resorting to the K\"ahler resolved picture, where only one side of the triangle (i.e. one leg of the web) can be deformed at a time. With those methods, it was impossible to decide whether deformations could be implemented simultaneously on all three legs. 
Because we have taken a different approach here, we see that the deformations can be superposed, just as in the previous section. The reason we can circumvent the limitations of \cite{Bourget:2023wlb} is the following: In that paper, it was crucial to perform a global projection of the M-theory geometry to IIA string theory, using the technology developed by \cite{Closset:2018bjz, Benini:2009qs}. However, for such a projection to work, only one side of the toric polytope can be deformed at a time. In addition, it is necessary to at least partially blow-up the threefold, in order to get a sensible weakly coupled QFT in 5d, where one can understand the physics of the IIA branes wrapping compact cycles. In this paper, we circumvent both needs in one fell swoop by working directly with the toric ideals describing patches of the singular CY threefold.

Hence, we can claim that the following deformations for $T_N$ are the most general. They are computed stepwise as follows: 
\begin{itemize}
    \item First, work on one leg, for instance
\begin{equation}
    X_1 X_2 X_3 = W^N \,, \qquad X_1 X_2 X_3 = \prod_{\lambda_i}(W^{\lambda_i} +\alpha_i X_1) \,,
\end{equation}
whereby $[\lambda_1, \ldots, \lambda_k]$ gives a partition of $N$, for the patch where $X_1 \neq 0$.
\item Now expand all the terms in powers of $W$ and extract all deformation terms (i.e. everything but the $W^N$ term).
\item Repeat for each leg by systematically replacing $W^N$ with such a product.
\item Add all these deformations to $W^N$.
\end{itemize}
We claim that such deformations can be safely superposed\footnote{We thank Julius Grimminger and Guillermo Arias-Tamargo for pointing out that the s-rule does not restrict the deformations of the $T_N$ geometries.} For example, for $T_3$, suppose we want our three legs to have the following partitions: $\Lambda_1 = [1,1,1], \Lambda_2 = [2,1], \Lambda_3 = [3]$. Then we simply work out
\[W^3 \mapsto \prod_{i=1}^3 (W-\alpha_i X_1) = \sum_{i=1}^3 (-1)^i\,W^{3-i} \,X_1^i\,e_i(\vec{\alpha})\]
for leg 1, then
\[W^3 \mapsto (W^2-\beta_1 X_2)\,(W-\beta_2 X_2) \]
for leg 2, and finally
\[W^3 \mapsto W^3 - \gamma X_3\]
for the third leg.
Now we extract all the deformations and superpose them into the following hypersurface:
\begin{equation}
W^3-\Big[e_1(\vec{\alpha})\,X_1+\beta_2 X_2\Big]W^2
\;+\;
\Big[e_2(\vec{\alpha})\,X_1^2-\beta_1 X_2\Big]W
\;+\;
\Big[\beta_1\beta_2\,X_2^2-e_3(\vec{\alpha})\,X_1^3-\gamma X_3\Big]
\end{equation}

% where
\[
e_1(\vec{\alpha})=\alpha_1+\alpha_2+\alpha_3,\qquad
e_2(\vec{\alpha})=\alpha_1\alpha_2+\alpha_1\alpha_3+\alpha_2\alpha_3,\qquad
e_3(\vec{\alpha})=\alpha_1\alpha_2\alpha_3
\]
Our claim is that this is the deformation corresponding to figure \ref{fig:T3}.

\section*{Open questions}
In this paper, we gave an algebro-geometric diagnostic for s-rule violations. This requires further exploration and testing in more complicated setups. Most vexing is the issue explained in the paper, that algebraic-geometry seems to treat anti-branes as poles, in accordance to how divisors are related to homology as
\[ {\rm Div}(f) = [{\rm zeroes}(f)]-[{\rm poles}(f)]\,.\]
This picture, albeit suggestive, requires a firm physical explanation. While it is acceptable to subtract anti-branes from branes at the level of B-branes, M-theory, being inherently a backreacted account of IIA with D6-branes, abhors this simplification. It is nevertheless intriguing that this works heuristically.

\section*{Acknowledgments}
I would like to thank Emilie Despontin, Mario De Marco, Roberto Valandro, Michele Del Zotto, Julius Grimminger and Guillermo Arias-Tamargo for many useful discussions. I would also like to thank Francesco Benini for initial stage collaboration.
The research of A.C. is funded through an ARC advanced project, and further supported by IISN-Belgium (convention 4.4503.15).
A.C. is a Senior Research Associate of the F.R.S.-FNRS (Belgium).

\section*{Appendix: The Schur Complement as a Homological Quotient} \label{app:schur}

In the framework of tachyon condensation, D-branes are described as objects in the derived category of coherent sheaves. Specifically, a D-brane bound state is represented by a two-term chain complex $E \xrightarrow{\mathcal{T}} F$, and its physical degrees of freedom are captured by the mapping cone of the tachyon field $\mathcal{T}$.

Consider a generic brane system represented by the direct sum complex:
\[
E_1 \oplus E_2 \xrightarrow{\qquad \mathcal{T} \qquad} F_1 \oplus F_2
\]
where the tachyon map has the block form:
\[
\mathcal{T} = \begin{pmatrix} T_1 & \tilde{Q} \\ Q & T_2 \end{pmatrix}
\]
Suppose the sub-system $(E_2, F_2)$ undergoes complete perturbative annihilation. Mathematically, this implies the localized tachyon field $T_2$ is an isomorphism.

To extract the surviving IR physics, we first perform an algebraic preparation. We can decouple the system by performing Gaussian block-elimination on $\mathcal{T}$. This corresponds to applying invertible basis changes (automorphisms) to the target and source bundles:
\[
\begin{pmatrix} \mathbf{1}_{F_1} & -\tilde{Q}\cdot T_2^{-1} \\ 0 & \mathbf{1}_{F_2} \end{pmatrix} 
\begin{pmatrix} T_1 & \tilde{Q} \\ Q & T_2 \end{pmatrix} 
\begin{pmatrix} \mathbf{1}_{E_1} & 0 \\ -T_2^{-1}\cdot  Q & \mathbf{1}_{E_2} \end{pmatrix} 
= \begin{pmatrix} T_1 - \tilde{Q}\cdot T_2^{-1} \cdot Q & 0 \\ 0 & T_2 \end{pmatrix}
\]
Because these row and column operations are strict automorphisms, the new complex is quasi-isomorphic to the original one. The system has explicitly block-diagonalized into a direct sum of two independent chain complexes:
\[
\Big( E_1 \xrightarrow{\; S \;} F_1 \Big) \;\oplus\; \Big( E_2 \xrightarrow{\; T_2 \;} F_2 \Big)
\]
where $S = T_1 - \tilde{Q} \cdot T_2^{-1} \cdot Q$ is the Schur complement. 

We now interpret this result homologically. The physical D-brane bound state is the mapping cone of this total complex. Because the mapping cone of a direct sum is the direct sum of the individual cones, we have:
\[
\text{Cone}(\mathcal{T}) \cong \text{Cone}(S) \oplus \text{Cone}(T_2)
\]
Since $T_2$ is an isomorphism, the complex $E_2 \xrightarrow{T_2} F_2$ is exact. Its mapping cone is purely acyclic, carrying zero cohomology:
\[
\text{Cone}(T_2) \cong 0
\]
Physically, this represents the perturbative vacuum of the annihilated brane pair. Stripping this trivial block away leaves us with the true IR-effective degrees of freedom:
\[
\text{Cone}(\mathcal{T}) \cong \text{Cone}(S)
\]
This demonstrates that if a localized sub-brane completely condenses, homologically quotienting it out of the system is rigorously equivalent to evaluating the Schur complement of the tachyon block matrix.

%%%%%%%%%%%%%%%%%%%%%%%%%%%%%%%%%%%%%%%%%%%%%%%%%%%%%%%%%%%%%%%%%%%%%%
% === HARD STOP: DO NOT MODIFY BELOW THIS LINE ===
%%%%%%%%%%%%%%%%%%%%%%%%%%%%%%%%%%%%%%%%%%%%%%%%%%%%%%%%%%%%%%%%%%%%%%
\newpage
\phantomsection % Fixes hyperref anchoring for the bibliography
\addcontentsline{toc}{section}{References} % Forces it into your Overleaf outline

\bibliographystyle{JHEP}
\bibliography{biblio}

\end{document}